# Open Shouldn't Mean Exempt: Open-Source Exceptionalism and Generative AI

David Atkinson[1,2]

## Abstract

Open-source status should not shield generative artificial intelligence systems from ethical or legal accountability. Through a rigorous analysis of regulatory, legal, and policy frameworks, this Article contends that open-source GenAI must be held to the same standards as proprietary systems. While recognizing the value of openness for scientific advancement, I propose a narrowly tailored safe harbor for bona fide, non-commercial research, conditioned on strict compliance with defined criteria. This Article critically examines and refutes the core claims of open-source exceptionalism—namely, that open-source GenAI disrupts entrenched oligopolies, democratizes access, and uniquely drives innovation. The evidence shows that open-source GenAI can facilitate unlawful conduct, exacerbate environmental harms, and reinforce existing power structures. Rhetoric around "democratization" and "innovation" often serves as an unsubstantiated basis for regulatory exemptions not afforded to proprietary systems. This Article ultimately advocates for a framework that promotes responsible AI development, balancing openness with robust legal and ethical safeguards and a clear-eyed assessment of societal impacts.

[1] Assistant Professor of Instruction, Business, Government and Society Department, McCombs School of Business, University of Texas, Austin; Law Fellow, Georgetown Law

[2] This paper uses the Allen Institute for Artificial Intelligence (Ai2) as a representative example throughout. I was in-house counsel at Ai2 from June 2023 until January 2025, where I worked on related matters. Everything shared in this paper is already publicly available.

# I. Introduction

In the evolving landscape of artificial intelligence, numerous well-funded entities across both nonprofit and for-profit sectors claim to offer open-source generative artificial intelligence (GenAI) tools, including chatbots and image generators. These entities, ranging from research institutes to technology conglomerates,[3] assert that releasing datasets, models, and code with minimal restrictions advances the public good and democratizes access to cutting-edge AI technologies. Yet, the precise contours of what constitutes "open-source" GenAI remain deeply contested, and the implications of these claims warrant careful scrutiny.[4]

To those outside the AI space, open-source projects may seem arcane. However, they function similarly to proprietary models and, in many ways, are poised to impact society broadly, regardless of whether individuals have a direct interest in or interaction with the technology. Unfortunately, these entities often engage in the same potentially unlawful actions and contribute to the same environmental harms as the closed-source entities they seek to distinguish themselves from. Indeed, proponents of open-source GenAI often frame these entities, or at least their open-source contributions, as indispensable champions of the public good. Advocates such as EleutherAI and Ai2 have publicly characterized open source as a universally positive force, often prioritizing openness over responsible or ethical AI development. However, this position warrants scrutiny in light of documented legal and ethical challenges.[5]

[3] 01.AI (Yi), Alibaba (Qwen), Allen Institute for Artificial Intelligence (OLMo), Common Crawl (Common Crawl crawled datasets), Databricks (DBRX), EleutherAI (The Pile), Google (Gemma), High-Flyer Capital Management (DeepSeek-R1), Hugging Face (SmolLM, FineWeb), Kaggle (various datasets), Meta (Llama), Mistral (Mixtral), StabilityAI (Stable Diffusion, Beluga), Technology Innovation Institute (Falcon), xAI (Grok), and various academic contributors like Stanford's RedPajama dataset.

[4] I'll use the word "artifact" to refer to the components of AI systems, like the datasets, code, evaluations, and so on.

This Article challenges that presumption of innate goodness. The artifacts these entities produce can be divided into two categories: mostly open and mostly closed.[6] Based on the criteria this paper later addresses, the "mostly open" camp includes Ai2, EleutherAI, and Hugging Face. The "mostly closed" camp includes 01.AI, Alibaba, Databricks, Google, Meta, Mistral, Stability AI, the Technology Innovation Institute, and xAI. For analytical purposes, this paper employs the names of the most vocal and influential entities from each camp to represent their respective approaches: the nonprofit Ai2, the for-profit Hugging Face, and the for-profit Meta.[7]

Open-source development can be a powerful engine for progress. At a conceptual level, it enables freer exchange of ideas, supports the interrogation and reproduction of findings, and facilitates broader review and improvement of code and data than private companies typically achieve. These features enable faster iteration, more diverse experimentation, and a deeper understanding of how technology works.

This Article advances the thesis that the purported benefits of open-source GenAI do not justify exempting these systems from established legal and ethical standards. Acts such as copyright infringement, privacy violations, and contractual breaches are not rendered permissible merely because of openness. Furthermore, there is scant empirical evidence that open-source GenAI delivers greater societal benefit than its proprietary counterparts, undermining arguments for special regulatory treatment. In fact, such practices sometimes cause demonstrable harm, and the open-source GenAI community has failed to provide an adequate response. Instead, ethical and legal concerns are baked into open models that are then freely shared for anyone to use for any purpose, ranging from disinformation[8] and nonconsensual intimate image generation to the proliferation[9] of child sexual abuse material,[10] and are often dismissed as mere byproducts of technological immaturity.[11]

---

[5] *E.g.*, allenai.org: "More than open," "We are open first," and the mouthful "Truly open breakthrough AI"

[6] See Sec. VII(A) *infra*

[7] *E.g.,* Devin Coldewey, *Is Open Source AI Even Possible, Let Alone the Future? Find Out at Disrupt 2024*, TechCrunch (Aug. 27, 2024), https://techcrunch.com/2024/08/27/is-open-source-ai-even-possible-let-alone-the-future-find-out-at-disrupt-2024/ (last visited July 7, 2025).(Ai2, Hugging Face, and Meta invited to discuss open-source at one of the most well-known technology conferences in the world, TechCrunch Disrupt)

[8] *E.g.*, Angela Yang, *AI Image Misinformation Surged, Google Research Finds*, nbcnews.com (last visited July 7, 2025), https://www.nbcnews.com/tech/tech-news/ai-image-misinformation-surged-google-research-finds-rcna154333.

Open-source advocates argue that these problems would diminish as more open-source GenAI becomes available for research and experimentation. This circular reasoning overlooks that leading supporters of open-source GenAI systematically undervalue the risk that malicious actors will intentionally misuse these systems to produce harmful outputs. When acknowledged, such risks are dismissed as necessary trade-offs for advancing knowledge.

Contrary to their self-presentation as champions of the public good, open-source GenAI often facilitates copyright infringement and environmental harm without disrupting the oligopolistic status quo. Supporters advocate "openness" as an inherently worthy goal. This reductive framing obscures complex considerations. Openness offers an alluringly simple metric that is relatively easy to measure and intuitively appealing. Other arguments for open source center on several core claims: environmental benefits, marketplace improvements, technological progress as an inherent good, and the democratization of a technology that advocates compare to the discovery of fire or electricity. This paper will systematically explore and challenge each of these claims.

To be sure, there is nothing inherently wrong with open-source GenAI. When pursued responsibly, open-source GenAI offers unique avenues to advance scientific understanding of GenAI, foster innovation, and enable more equitable access to the GenAI supply chain. The problem arises when entities producing widely used GenAI artifacts fail to adhere to ethical and responsible model development practices. Rather than demonstrating that adequate models can be built through innovation and advances in fundamental research, these entities measure themselves against for-profit companies (or are themselves for-profit companies) driven by incentives to cut corners. This creates competition for first-mover advantage, economies of scale, brand recognition, and network effects that entrench market position. Such dynamics foster a culture incompatible with responsible development and create relentless pressure to demonstrate exponential growth in users or revenue. These incentives directly conflict with goals such as environmental sustainability, equitable benefit distribution, and the resolution of complex societal challenges that technological progress alone cannot address.

---

[9] *E.g.,* Natasha Singer*, Nudes, Deepfakes, and AI: Inside One High School's Fight Against Online Harassment*, N.Y. Times (Apr. 8, 2024), https://www.nytimes.com/2024/04/08/technology/deepfake-ai-nudes-westfield-high-school.html.";Bilva Chandra, *Analyzing Harms from AI-Generated Images and Safeguarding Online Authenticity*, RAND Corp. (2024), https://www.rand.org/content/dam/rand/pubs/perspectives/PEA3100/PEA3131-1/RAND_PEA3131-1.pdf.

[10] *E.g.,* Matteo Wong, *AI Is Triggering a Child Sex Abuse Crisis, The Atlantic (Sept. 18, 2024),* https://www.theatlantic.com/newsletters/archive/2024/09/ai-is-triggering-a-child-sex-abuse-crisis/680053/.; David Thiel, *Investigation Finds AI Image Generation Models Trained on Child Abuse*, Stan. Cyber Pol'y Ctr., https://cyber.fsi.stanford.edu/news/investigation-finds-ai-image-generation-models-trained-child-abuse (last visited July 7, 2025).

[11] Ali Fahardi, CEO of Ai2: "To me, we're having all of these conversations because the technology stack is just not ready yet." Video, *The Increasing Support and Advantages of Open AI With Meta and Hugging Face*, TechCrunch (Sept. 15, 2023), https://techcrunch.com/video/the-increasing-support-and-advantages-of-open-ai-with-meta-and-hugging-face/. at 19:31. He then pivots back to how openness can lead to more progress in understanding AI and improvements in the capabilities of AI. In other words, any harm caused by open source is merely a technological problem that technological progress can solve.

Despite these tensions, many open-source GenAI advocates maintain that their work warrants special treatment under existing law.[12] Granting such exemptions would be both unjust and unwise, signaling that society values certain technological artifacts so highly that it will disregard principles of copyright law, contract law, and environmental stewardship. This Article argues that open-source GenAI developers must adhere to the same legal and ethical standards as their proprietary counterparts.

Nevertheless, this Article recognizes that narrowly tailored exemptions may be warranted for genuine scientific research conducted by nonprofit academic institutions. To qualify for such a safe harbor, an institution must: (i) operate on a nonprofit basis; (ii) engage primarily in scientific research rather than product development; (iii) release artifacts solely for non-commercial use under share-alike licenses; and (iv) restrict access to verified researchers affiliated with nonprofit institutions of higher education, or acting under their direction, for scholarly research and teaching. These safeguards would not unduly constrain fundamental research but would instead bolster existing legal frameworks that promote the creation and sharing of intellectual property, facilitate robust exchange of ideas, and stimulate economic growth.

Finally, a note to clarify what this paper is not. First, this is not an argument in favor of closed-source GenAI. Rather, it advocates for responsibly developed GenAI systems that are both safe and ethical. Second, this is not an analysis of AI in general but a targeted exploration of issues unique to GenAI. Third, the legal discussion is intended to illustrate how the law applies to open-source GenAI, just as it does to proprietary systems. It is not meant to be an exhaustive indictment of every way open-source GenAI may violate the law. Fourth, this paper largely avoids distinguishing between claims that apply to model inputs and those that apply to outputs. The fact that an action could be illegal regardless of when it occurs is sufficient for this paper's argument. Finally, while this paper primarily focuses on GenAI model developers, it also applies to individuals who build upon open-source GenAI where applicable.

To understand the implications of open-source GenAI, it is essential to first clarify what "open source" means in the context of artificial intelligence and how these principles apply to generative models.

# II. Open Source and GenAI

The term "open source" in software emerged in the 1990s, revolutionizing how technology is developed and shared. Open source enables projects to incorporate contributions from anyone in the world willing to participate. Some of the most impactful open-source projects include frameworks that power AI development (TensorFlow, PyTorch), widely used programming languages (Python, Go, Swift), web browsers (Firefox, Chromium), coding environments (VS Code), and even cryptocurrencies like Bitcoin.

[12] This paper will describe specific examples below, but they include Meta arguing in litigation that its open-weights models are socially beneficial as a way to avoid copyright infringement claims, the EU AI Act exempting transparency requirements for open-source AI models, and Stanford researcher Fei-Fei Li claiming that some provisions in California's 2024 AI bill would "stifle" and "devastate the open-source community," Fei-Fei Li, *'The Godmother of AI' says California's well-intended AI bill will harm the U.S. ecosystem*, YAHOO FIN. (Aug. 6, 2024), https://finance.yahoo.com/news/godmother-ai-says-california-well-123044108.html.), and others.

The Open Source Initiative (OSI), the leading authority on open-source standards, outlines ten criteria that software licenses must meet to qualify as "open source." These include principles such as free redistribution, access to source code, the right to create derivative works, and non-discrimination against persons or fields of endeavor.[13]

While these principles have long guided open-source software development, the advent of AI introduces new complexities. Unlike conventional software, AI systems consist of multiple interdependent components: training data, training code, and model parameters (often called "weights and biases"). In response, the Open Source Initiative (OSI) has advanced a working definition for Open Source AI that attempts to address these unique challenges:

> An Open Source AI is an AI system made available under terms and in a way that grants the freedoms to:
>
> - *Use* the system for any purpose and without having to ask for permission.
> - *Study* how the system works and inspect its components.
> - *Modify* the system for any purpose, including to change its output.
> - *Share* the system for others to use with or without modifications, for any purpose.
>
> These freedoms apply both to a fully functional system and to discrete elements of a system. A precondition to exercising these freedoms is to have access to the preferred form to make modifications to the system.[14]

The complications begin with data requirements. The most contentious debates within the open-source community concern the OSI's approach to data transparency, which allows companies such as Meta to assert open-source status even while withholding the underlying training data for their models.[15] Notably, OSI's "Open Source AI" definition does not require entities to fully disclose their data, as they must with code. Instead, it mandates only "sufficiently detailed information":

> - Data Information: Sufficiently detailed information about the data used to train the system so that a skilled person can build a substantially equivalent system. Data Information shall be made available under open-source-approved terms.
>     - In particular, this must include: (1) the complete description of all data used for training, including (if used) of unshareable data, disclosing the provenance of the data, its scope and characteristics, how the data was obtained and selected, the labeling procedures, and data processing and

[13] *The Open Source Definition*, Open Source Initiative, https://opensource.org/open-sourced

[14] Open Source Initiative, *The Open Source AI Definition*, https://opensource.org/ai/open-source-ai-definition

[15] *Open Source AI Available to All, Not Just the Few*, Meta (June 16, 2023), https://www.facebook.com/Meta/videos/open-source-ai-available-to-all-not-just-the-few/472483165627318/.

> filtering methodologies; (2) a listing of all publicly available training data and where to obtain it; and (3) a listing of all training data obtainable from third parties and where to obtain it, including for fee.[16]

The OSI's requirement for only "sufficiently detailed information" about training data falls short of enabling true replication of most GenAI systems. Scientific replication requires reproducing results by precisely replicating the actions performed on identical artifacts. Under current definitions, researchers are typically limited to emulating rather than replicating existing systems—producing functionally similar but not identical models.[17] This nuance reveals that openness in AI exists on a spectrum rather than as a binary. As researchers Andreas Liesenfeld and Mark Dingemanse observe, "ubiquity and free availability are not equal to openness and transparency."[18] Meta's approach exemplifies the gradations of openness in AI, where some elements may be shared, but full reproducibility, a cornerstone of the scientific method, remains elusive. This paper will explore these components and their varying degrees of openness in more detail below.

## A. Benefits of Open-Source GenAI for GenAI Researchers

This section focuses exclusively on research-related benefits that enable researchers to examine models and understand how open-source GenAI artifacts function. Proponents of legal exemptions for open-source GenAI often highlight contributions to scientific advancement through open experimentation, noting that exemptions for scientific research sometimes exist where open-source exemptions do not. Because open-source artifacts are designed for broad availability, advocates overstate research-based merits to sidestep legal issues, including mass copyright infringement, that affect both open and closed-source systems. In other words, open-source entities tend to focus heavily on a commitment to scientific research as a justification for exemption from laws that would otherwise apply to them.

Given this context, here are compelling reasons why researchers might prefer deploying open-source models on local machines (i.e., on servers not controlled or accessible by GenAI companies) rather than relying solely on API access[19], as is common with systems like ChatGPT, Gemini, and Claude:[20]

### 1. Experimental Stability and Control

Running a model locally helps ensure the system under study remains stable from one day to the next. In contrast, companies like OpenAI or Anthropic may introduce tweaks or updates without warning, a phenomenon known as "model drift." A prominent example was when OpenAI rolled back an update that made their system overly sycophantic.[21] Such unannounced changes undermine experimental control; when the artifact under study is in constant flux, it becomes impossible to definitively attribute observed outcomes to the experimental conditions.

---

[16] Open Source Initiative, *The Open Source AI Definition*, https://opensource.org/ai/open-source-ai-definition

[17] Data is not the only limiting factor for reproducibility of models. For example, it'd be all but impossible to exactly replicate the randomized weights when training is initiated unless it is provided by the originator.

[18] Abeba Birhane, Vinay Prabhu, Sang Han, Vishnu Naresh Boddeti & Alexandra Sasha Luccioni, *Rethinking open source generative AI*, FOUN. & TRENDS PRIV. & SEC., Vol. 1, No. 2, pp. 1-38 (June 5, 2024), https://dl.acm.org/doi/10.1145/3630106.3659005.

[19] An API connection for GenAI is a standardized way for researchers to send requests to and receive information from the GenAI model(s), allowing it to access real-time data or perform external actions. It is a more scalable and direct form of access to the models rather than a chat window intermediating the exchange.

[20] A special thanks to Jacob Morrison from the Allen Institute for Artificial Intelligence for identifying these benefits.

## 2. Research Continuity and Reproducibility

Closed-source providers can and do deprecate models or discontinue support altogether. For researchers conducting long-term studies on specific models, losing access midway can derail research efforts and make replicating results impossible.

## 3. Avoiding Data Contamination

A common research technique involves administering benchmark questions to models and analyzing responses. A significant drawback of proprietary systems is that providers may retain and train on submitted questions. This practice, known as "data contamination," erodes evaluation integrity, as models may "learn the test" rather than demonstrating genuine capabilities.[22] Conversely, with open-source models run locally, the developers have no access to the test questions, ensuring the model's performance on well-constructed benchmarks remains statistically valid over time.

## 4. Cost-Effective Customization

Open-source systems enable researchers to fine-tune models more cost-effectively than developing new models from scratch. Fine-tuning existing open-source models is relatively inexpensive (requiring less compute, electricity, and time) and allows extensive customization. API-based models, by contrast, often impose strict limits on customization, making it difficult to isolate and measure the direct impact of fine-tuning on performance.

## 5. Mechanistic Interpretability

Perhaps the most significant advantage of open-source models is access to the underlying model weights. This transparency enables researchers to conduct in-depth interpretability studies, for example, by identifying which artificial neurons activate in response to specific inputs. This level of mechanistic insight is unattainable with closed-source "black box" models, where researchers are limited to asking superficial questions about a model's overall capabilities.

In summary, open-source models empower researchers to explore not only what models can do but also how and why they behave in a particular way. Such insight is essential for rigorous scientific inquiry and remains beyond the reach of API-based, closed systems.

The preceding section has demonstrated that open-source GenAI offers researchers significant advantages, including experimental control, reproducibility, and interpretability. However, the impact of open-source

---

[21] OpenAI, *Sycophancy in GPT-4o: what happened and what we're doing about it*, https://openai.com/index/sycophancy-in-gpt-4o/

[22] Alex Reisner, *Chatbots Are Cheating on Their Benchmark Tests*, https://www.theatlantic.com/technology/archive/2025/03/chatbots-benchmark-tests/681929/

GenAI extends beyond the research community. The following section explores how openness benefits a broader range of stakeholders, including businesses and the general public, through increased transparency, extensibility, and data control.

# III. Additional Benefits of Open-Source GenAI

The benefits of open-source GenAI extend beyond research applications. In fact, most open-source developers likely rely on a user base composed chiefly of non-research users.[23] The broader benefits often mirror those of traditional open-source software, including transparency, extensibility, and data secrecy, all while permitting commercial use.

## A. Transparency

Transparency enables "looking under the hood." For GenAI, this primarily requires access to key components, such as model weights and training data. Although the "transparent" model weights themselves remain inscrutable to most people, the datasets offer tangible insight. Such access can empower stakeholders, including copyright holders, privacy advocates, and regulatory bodies, to participate in informed discussions about a model's ethical, legal, and social implications, since model knowledge and capabilities are limited to what they are trained on.

However, there are reasons to be skeptical that transparency alone will have the transformative impact that open-source boosters claim. For example, Axios uncritically quoted Ai2's CEO, Ali Farhadi, in noting that "Farhadi said that AI makers won't be able to earn back the public's trust until they can understand how their models produce a particular output — and they won't be able to do that until their data is fully available to researchers."[24]

Evidence does not support this position. Few members of the public understand how electricity, car engines, mobile phones, or internet protocols work, yet society adopted these technologies readily and continues using them without hesitation. Focusing on "transparency" as a panacea for GenAI's problems serves as a justification for advancing preferred policy positions and securing funding while avoiding the work of addressing harms created by technology, including the proliferation of misinformation, manipulation, and deepfakes.

## B. Extensibility and Modification

A key advantage of open-source models is their extensibility, meaning users can build on existing artifacts. Rather than developing new GenAI models from scratch, businesses can create an application "wrapper" (for example, a custom user interface) or fine-tune an open-source model for a specific application, such as chatbots for real estate websites or specialized coding assistants. Numerous companies already employ this strategy across diverse sectors, including law (e.g., Harvey), search (e.g., Perplexity), coding (e.g., Cursor), digital media (e.g., Adobe), and medical notetaking (e.g., Abridge). While many of these firms currently use closed models, the potential for deep modification is significantly broader with open-source alternatives.[25]

---

[23] If the focus were exclusively on researchers, one would expect to see more restrictive licenses that limit use to non-commercial or academic purposes only.

[24] Ali Farhadi, *AI Summit*, Axios (June 5, 2024), https://www.axios.com/2024/06/05/ali-farhadi-ai-summit?utm_source=newsletter&utm_medium=email&utm_campaign=newsletter_axioslogin&stream=top.

## C. Data Secrecy

Another benefit that appeals to businesses is enhanced control over their proprietary data. Although many GenAI companies offer enterprise-level subscriptions that promise data protection and assurances against using client inputs for further training, the most robust safeguard against information leakage is deploying "on-premise" (i.e., on company-owned servers) or self-hosted models (i.e., hosted on third-party servers). Open-source solutions enable on-premise deployment, keeping proprietary data and internal processes under company control.

While open-source GenAI offers tangible benefits in transparency and flexibility, these practical advantages are only part of the story. Many organizations also pursue open-source strategies for reputational, market, and regulatory reasons. Next, this paper examines these strategic motivations, revealing how openness is leveraged not only for technical progress but also for competitive and legal positioning.

# IV. Strategic Motivations for Embracing Open Source

While open-source GenAI principles emphasize scientific research and societal advancement, many companies promote open-source identities for strategic reasons, including reputation enhancement, market influence, and favorable legal and regulatory treatment.

## A. Reputation and Market Advantages

Companies benefit by positioning themselves as champions of an open, accessible approach to AI. They adopt slogans like Meta's "Available to all, not just the few"[26] or Hugging Face's mission to "democratize good machine learning, one commit at a time."[27] They also proclaim commitments to "Building breakthrough AI to solve the world's biggest problems"[28] with their "Truly open breakthrough AI" as with Ai2.[29] Such branding builds goodwill and drives greater adoption of their models and platforms, creating more opportunities for collaboration, funding, talent acquisition, and broader influence in the AI landscape.

---

[25] The lack of adoption by wrapper companies of open-source models also reveals how open-source models are not upsetting the tech giant oligopoly.

[26] *See* the series of Meta advertisements with this catch phrase: *Open-Source AI: Available to All, Not Just the Few*, Facebook (June 16, 2023), https://www.facebook.com/Meta/videos/open-source-ai-available-to-all-not-just-the-few/480468301772528/.; *Open-Source AI: Available to All, Not Just the Few*, Facebook (June 16, 2023), https://www.facebook.com/Meta/videos/open-source-ai-available-to-all-not-just-the-few/2145330219233115/.; *Open-Source AI: Available to All, Not Just the Few*, Facebook (June 16, 2023), https://www.facebook.com/Meta/videos/open-source-ai-available-to-all-not-just-the-few/2372423423093997/.

[27] Hugging Face, https://huggingface.co/huggingface.

[28] Allen Inst. for AI, About, https://allenai.org/about.

[29] https://allenai.org/

Each entity's strategic approach to open source is unique. For instance, Hugging Face provides a central hub for developers to discover, share, and collaborate on pre-trained models, datasets, and evaluation metrics. This community-driven strategy positions Hugging Face as an indispensable resource in the AI ecosystem. Active community participation creates network effects that enhance Hugging Face's ecosystem value and increase switching costs for users, similar to social media platforms. By positioning itself as the central hub for open-source AI, Hugging Face has established a dominant presence in the AI datasets and model-sharing market, reinforcing its influence and competitive advantage.[30] This positioning enables them to sell additional services and tools, further solidifying their business model and expanding their influence.

Meanwhile, Meta derives multiple benefits from providing the most downloaded open-weights models. By releasing models like Llama, Meta effectively crowdsources a massive amount of free research and development labor. The global community fine-tunes, evaluates, and builds upon their models, leading to rapid improvements, bug fixes, and novel applications that Meta might not have conceived of or executed as quickly independently.[31]

Notably, the goodwill of releasing models is so potent that even OpenAI, which has been famously not open for several years in almost any way, is slated to share an open-weight model in mid-2025.[32] This is despite OpenAI already being the most influential GenAI company as measured by annual recurring revenue.[33]

## B. Legal and Regulatory Advantages

Beyond reputational and market incentives, many companies pursue open-source strategies to gain legal and regulatory advantages. By aligning with open-source principles, entities may seek exemptions from certain statutory requirements, as illustrated by the EU AI Act and various privacy laws.

### The EU AI Act

The EU AI Act allows exemptions for "AI models that are released under a free and open source license, and whose parameters, including the weights, the information on the model architecture, and the information on model usage."[34] These exemptions include not having to comply with "transparency-related requirements."[35] Those transparency requirements are outlined in Article 50 and include the requirement that providers of AI systems designed to interact directly with individuals must inform users that they are dealing with an AI system unless it's obviously not an AI system. Similarly, open-source model providers are exempt from the obligation that providers of AI systems that create synthetic audio, images, video, or text (including general-purpose AI) must ensure these outputs are machine-readable and detectable as artificially generated or manipulated, using effective and reliable technical solutions.

---

[30] *E.g.*, As of September 2024, Hugging Face hosts over one million open AI models. https://the-decoder.com/hugging-face-is-growing-fast-with-users-creating-new-ai-repositories-every-10-seconds/

[31] Meta isn't known for innovation: it created Stories to copy Snapchat, Reels to copy TikTok, blue verification checkmarks to copy Twitter, Roll Call to copy BeReal, Threads to copy Twitter, disappearing messages to copy Snapchat, filters to copy Snapchat, etc.

[32] OpenAI's Open-Source AI Model Is Delayed, THE VERGE, https://www.theverge.com/news/685125/openais-open-source-ai-model-is-delayed

[33] ARR can be a proxy for how useful a product is. Presumably, people would not pay for something if they didn't find it valuable. The Information, AI-Native Startups Pass $15 Billion Annualized Revenue, https://www.theinformation.com/articles/ai-native-startups-pass-15-billion-annualized-revenue

[34] https://artificialintelligenceact.eu/recital/104/

[35] https://artificialintelligenceact.eu/recital/104/

The other primary exemption for open source is from the general requirement for providers of general-purpose AI models to maintain comprehensive and updated technical documentation for their models, including details on training, testing, and evaluation results. Additionally, those providers, but not open-source model providers, must create and make available information and documentation to developers who integrate their general-purpose AI models into other AI systems. The documentation must clearly explain the model's capabilities and limitations to help integrators comply with the AI Act.[36]

Notably, The Act's exemption "from compliance with the transparency-related requirements should not concern the obligation to produce a summary about the content used for model training and the obligation to put in place a policy to comply with Union copyright law, in particular to identify and comply with the reservation of rights pursuant to Article 4(3) of Directive (EU) 2019/790 of the European Parliament and of the Council."[37]

## US Copyright Law

This paper will focus solely on U.S. copyright law, both for simplicity and because it is likely the most consequential of all copyright laws globally.[38] There is no open-source exemption in U.S. copyright law. However, there is an affirmative defense to claims of copyright infringement for "fair use." The introductory paragraph to the fair use section states that "the fair use of a copyrighted work, including such use by reproduction in copies or phonorecords or by any other means specified by that section, for purposes such as criticism, comment, news reporting, teaching (including multiple copies for classroom use), *scholarship, or research*, is not an infringement of copyright." (emphasis added)[39]

As noted a few times already, one of the key stances of some open-source GenAI entities is that their models are for research purposes. Additionally, the first of the four fair use factors described in the statute requires consideration of the commercial nature of the use. While the commercial nature is obvious when people must pay to use a model, the issue becomes murkier when the model is free to use with the hope that the user will sign up for other services (as with Hugging Face) or when the free models are part of a for-profit platform (as with Meta).

[36] https://artificialintelligenceact.eu/article/53/
[37] https://artificialintelligenceact.eu/recital/104/
[38] The US produces most of the highest grossing films globally (https://www.boxofficemojo.com/year/world/2024/), the most streamed songs on Spotify (https://en.wikipedia.org/wiki/List_of_Spotify_streaming_records), the most bestselling books (https://lithub.com/these-were-the-bestselling-books-of-2024/), and so on.
[39] https://www.law.cornell.edu/uscode/text/17/107

### Privacy

A final type of legal exemption that entities may seek to lean on for open-source GenAI is from privacy laws. The two most relevant laws, due to their scale, are the General Data Protection Regulation (GDPR) under the EU and the California Privacy Rights Act (CCPA/CPRA).

The GDPR provides scientific research exemptions. "For the purposes of this Regulation, the processing of personal data for scientific research purposes should be interpreted in a broad manner including for example technological development and demonstration, fundamental research, applied research and privately funded research."[40] Assuming this criterion is satisfied,[41] and further assuming that the privacy rights granted to individuals "are likely to render impossible or seriously impair the achievement of the specific purposes, and such derogations are necessary for the fulfillment of those purposes,"[42] the research becomes exempt from the right of access by the data subject, the right to erasure ("right to be forgotten"), the right of rectification, the right to restriction of processing, and the right to object.[43]

The CPRA also has some notable exemptions. For one, it doesn't apply to non-profits like Ai2. For another, businesses "shall not be required to comply with a consumer's request to delete the consumer's personal information if it is reasonably necessary for the business, service provider, or contractor to maintain the consumer's personal information in order to…*Engage in public or peer-reviewed scientific, historical, or statistical research* that conforms or adheres to all other applicable ethics and privacy laws, when the business's deletion of the information is likely to render impossible or seriously impair the ability to complete such research, if the consumer has provided informed consent."[44] (emphasis added)

The discussion of strategic motivations has shown that open-source GenAI is often promoted as a means to gain market share and regulatory favor. However, these incentives must be weighed against the legal frameworks that govern AI development. The paper now turns to an analysis of why open-source GenAI should not be categorically exempt from existing laws, emphasizing the need for consistent legal accountability.

# V. Open Source and the Law: Why Open-Source GenAI Shouldn't Be Exempt

Although certain regulations, such as the EU AI Act, offer limited exemptions for open-source GenAI, extending such preferential treatment to other areas of law would be misguided. Granting categorical exemptions to open-source AI risks undermining foundational legal principles and eroding the rule of law.

---

[40] https://gdpr-info.eu/recitals/no-159/

[41] For example, is the entity truly engaged in fundamental research and experimenting to discover something scientifically, or is it mostly focused on just trying to replicate what closed-source companies have already done and releasing it for use by the masses?

[42] https://gdpr-info.eu/art-89-gdpr/

[43] https://gdpr-info.eu/art-89-gdpr/

[44] *The California Privacy Rights Act of 2020*, https://thecpra.org/

The following analysis builds on my previous scholarship.[45] Section V outlines only a few ways in which open-source datasets and models may violate the law, demonstrating that these concerns are not merely hypothetical. It is not meant to be exhaustive or to imply that *all* open-source datasets and models are unlawful.

This section will first dismantle the false equivalence between open source and scientific research before providing an overview of why open-source GenAI should not be automatically exempt from copyright law, contract law (via terms of service), licensing, or privacy law.[46] The unavoidable conclusion in each instance is that the law must apply with equal force to open-source and closed-source entities, and model developers who engage in unlawful conduct must be held liable regardless of their chosen distribution method.

## A. The False Equivalence of Open Source and Scientific Research

Open-source GenAI entities often cloak their arguments under the guise of scientific research, scholarship, and innovation. While these entities presumably genuinely believe their work contributes to societal progress, they may also strategically equate open source with scientific research because they understand that the law looks favorably on research. As shown above, this favorable treatment manifests in specific legal exemptions in the General Data Protection Regulation (GDPR)[47], the EU Copyright Directive[48], and the California Privacy Rights Act (CPRA)[49]. Similarly, exemptions under the EU AI Act specifically benefit open-source AI models.

It is likely not coincidental that in Ai2's "comment" to the U.S. Copyright Office's Notice of Inquiry on copyright and artificial intelligence, Ai2 mentions the word "research" 30 times. Ai2 focuses its justification for open-sourcing its AI artifacts solely on how it will benefit researchers, nonprofits, and governments:

- Open and transparent AI Models promote technical advancements that lead to more ethical models. Current technology gaps, such as the ability to watermark or remove personally identifiable

---

[45] David Atkinson, Jena Hwang, and Jacob Morrison, *Intentionally Unintentional: GenAI Exceptionalism and the First Amendment*, 23 First Amend. L. Rev. 173 (2025); David Atkinson, *Unfair Learning: GenAI Exceptionalism and Copyright Law* (forthcoming Chicago-Kent Intellectual Property Journal 2026); David Atkinson, *Putting GenAI on Notice: GenAI Exceptionalism and Contract Law*, NW. U. L. REV. ONLINE (2025)

[46] This paper does not discuss other important and relevant laws and regulations, such as torts or antitrust, or adjacent issues such as AI governance or national security.

[47] How GDPR Changes the Rules for Research, IAPP, https://iapp.org/news/a/how-gdpr-changes-the-rules-for-research (last visited July 7, 2025).

[48] Reed Smith, *AI in Entertainment and Media* (Feb. 2024), https://www.reedsmith.com/en/perspectives/ai-in-entertainment-and-media/2024/02/text-and-data-mining-in-eu.

[49] Gregory A. Gidus, *The Research Exception to the CCPA's Right to Deletion — Will It Ever Apply?*, Carlton Fields (2019), https://www.carltonfields.com/insights/publications/2019/research-exception-ccpa-right-deletion-apply#:~:text=Consultancies%20News%20Offices-,The%20Research%20Exception%20to%20the%20CCPA's%20Right%20to%20Deletion%20%E2%80%94%20Will,%20%20>*.

information, *can only be solved if researchers have access to the full details of AI Models*. Some of the ethical challenges that exist are technical problems that are yet to be solved (emphasis added)

- Openness is a prerequisite for external audits and scrutiny. In order to identify potential risks, create tools for harm mitigation, *and build AI literacy within the research community and the general public, AI researchers* need access to all Artifacts within an AI System for analysis (emphasis added)
- Reproducibility is a key tenet in scientific research. Open research has been a key component of the rapid progress in the development of AI technologies. *From academic departments to research labs*, the ability to replicate and build upon prior work has led to technical and economic advances. For example, the open release of BERT in 2018 led to a flurry of innovations in NLP4 (emphasis added)
- Equitable access. The training of AI Models is extremely computationally expensive, costing millions of dollars to complete. This creates an unbalanced concentration of power where only a small handful of companies and institutions have the resources to undergo the process, which in turn gives those companies exclusive access to data and research avenues. By making our OLMo model openly available, *we hope to empower academic institutions, government agencies, and other nonprofit organizations* with access to a state-of-the-art AI Model (emphasis added)[50]

The conclusion of their comment states that:

> AI is developing and improving at an astounding pace, and numerous AI Models and AI Systems already exist that will be affected by any decisions made by the Copyright Office, Congress, or the Courts. AI has tremendous potential to improve lives, and it also poses risks and harms. We contend that an important way to reduce the latter is *through increased scientific research* which includes access to robust and diverse data sources. We believe our recommendations allow for the flourishing of AI innovation while balancing the purpose of copyright law and the needs of copyright owners. *As a research-first nonprofit institute* that believes in an ethical, interdisciplinary, science-focused approach free from any profit motive, we appreciate the opportunity to share our thoughts and extend an offer to be of future assistance, as needed.[51] (emphasis added)[52]

Hugging Face performs a similar obfuscation. Their opening paragraph states that "The following comments are informed by our experiences as an open platform for state-of-the-art (SotA) AI systems, *working to make AI accessible and broadly available to researchers* for responsible development." (emphasis added).[53]

---

[50] Allen Institute for Artificial Intelligence, Response to the U.S. Copyright Office, Library of Congress Notice of inquiry and request for comments (RFC) (Nov 1, 2023) https://www.regulations.gov/comment/COLC-2023-0006-8762

[51] Allen Institute for Artificial Intelligence, Response to the U.S. Copyright Office, Library of Congress Notice of inquiry and request for comments (RFC) (Nov 1, 2023) https://www.regulations.gov/comment/COLC-2023-0006-8762

[52] Allen Institute for Artificial Intelligence, Response to the U.S. Copyright Office, Library of Congress Notice of inquiry and request for comments (RFC) (Nov 1, 2023) https://www.regulations.gov/comment/COLC-2023-0006-8762

[53] Hugging Face, Response to the U.S. Copyright Office, Library of Congress Notice of inquiry and request for comments (RFC) (Nov 1, 2023) https://www.regulations.gov/comment/COLC-2023-0006-8969

They go on to favorably quote researchers who championed the open-source model GPT-Neo, which was trained on the freely shared Books3 dataset of nearly 200,000 books protected by copyright: "Our research would not have been possible without EleutherAI's complete public release of The Pile dataset and their GPT-Neo family of models." In other words, sharing those copyrighted books with everyone is a *good* thing in Hugging Face's view.[54]

However, equating openness and scientific research is a fallacy. Open-sourcing a model does not inherently advance scientific inquiry, nor does widespread adoption equate to research-driven use. The over one billion downloads of Llama are not driven by a billion people who are suddenly and deeply interested in scientific research. Moreover, if an activity is otherwise unlawful, making it open source does not render it lawful. Scientific research neither requires nor is defined by open-source distribution.

More importantly, even when the stated goal *is* research, the method of release matters. For example, Ai2 does not need to release models and datasets under a permissive commercial license, such as Apache 2.0 or ODC-BY,[55] to promote scientific progress; it could use a more restrictive license limited to non-commercial research. If the true purpose were scientific advancement, restrictive licensing would better serve that goal while preventing commercial exploitation.

Depending on the training data or potential for malicious use, open-source artifacts may be highly detrimental to society. A flat exemption or even leniency for open-source could allow entities to engage in otherwise illegal activities, such as releasing datasets of pirated movies, songs, and books as open-source under the guise of scientific research. This creates a regulatory loophole that undermines the very purposes for which these exemptions were designed.

Privacy laws like GDPR and CCPA/CPRA create exemptions for nonprofits or scientific research with the understanding that data will be used exclusively for those purposes. The logic of these exemptions collapses when entities aggregate and clean vast amounts of data (including material scraped in violation of terms of service or copyright, or obtained from pirate sites) and then release it for unrestricted use by any party, including for-profit companies. If entities like Ai2 can collect and redistribute data under exemptions unavailable to for-profit entities, the exemption becomes a loophole. For-profit entities need only wait for Ai2 to perform actions they could not legally perform themselves, then download the resulting data, code, and models.

Similar controversy surrounds the EU AI Act, which currently lacks a clear definition of open source. Meta has marketed its Llama models as open source and appears positioned to claim open-source exemptions under the Act. Meta is not alone. In their coalition paper regarding the EU AI Act, GitHub, EleutherAI, Hugging Face, LAION, and others characterize "Open foundation models like those developed by EleutherAI, LAION or BigScience's BLOOM" as "*first and foremost research artifacts*." [56] (emphasis added). Yet none of their most popular artifacts are limited to research purposes, revealing the strategic nature of the "research" framing.

---

[54] Hugging Face, Response to the U.S. Copyright Office, Library of Congress Notice of inquiry and request for comments (RFC) (Nov 1, 2023) https://www.regulations.gov/comment/COLC-2023-0006-8969

[55] This paper will explain these licenses in more detail below.

[56] *Supporting Open Source and Open Science in the EU AI Act*, GitHub Blog (July 2023), https://github.blog/wp-content/uploads/2023/07/Supporting-Open-Source-and-Open-Science-in-the-EU-AI-Act.pdf.

An activity that is otherwise unlawful does not become lawful merely because it is labeled "open source" or "for research." Scientific research does not require breaking the law, and open source does not require unrestricted commercial use.

## B. Copyright

Copyright law is implicated at every stage of the GenAI lifecycle. Open-source entities are no more inherently compliant than proprietary counterparts.[57] These entities reproduce vast amounts of copyrighted material without authorization through scraping or by copying datasets others have compiled, distribute these works by packaging them into datasets, train and release models capable of generating infringing derivative works, and distribute these models with minimal use restrictions. Moreover, entities like Hugging Face, which hosts the datasets and models that are trained on these datasets and make them freely and easily downloadable, are often aware of legally suspect datasets (e.g., ones that likely contain at least thousands of copyrighted works without copyright owner authorization)[58] but take no action to remove the datasets from their platform.

### The Fair Use Analysis

Ultimately, many open-source GenAI entities invoke the doctrine of fair use as a defense against copyright infringement claims. However, the application of fair use in the context of large-scale data scraping and model training is unsettled and highly fact-dependent, requiring careful consideration of relevant precedent and the statutory factors:

1. the purpose and character of the use, including whether such use is of a commercial nature or is for nonprofit educational purposes;
2. the nature of the copyrighted work;
3. the amount and substantiality of the portion used in relation to the copyrighted work as a whole; and
4. the effect of the use upon the potential market for or value of the copyrighted work. [59]

If the factors, when weighed together, favor fair use, then there is no copyright infringement.

---

[57] Interestingly, prominent legal scholars such as Mark Lemley and James Grimmelmann taken an opposing view. Ars Technica, *Study: Meta's Llama 3.1 Can Recall 42 Percent of the First Harry Potter Book*, Ars Technica (June 2025), https://arstechnica.com/features/2025/06/study-metas-llama-3-1-can-recall-42-percent-of-the-first-harry-potter-book/./ ("There's a degree to which being open and sharing weights is a kind of public service," Grimmelmann told me. "I could honestly see judges being less skeptical of Meta and others who provide open-weight models.")

[58] This does not include other legal issues, like Hugging Face hosting "5000 models used to generate nonconsensual sexual content of real people." Emanuel Maiberg, *Hugging Face Is Hosting 5,000 Nonconsensual AI Models of Real People*, 404 Media, https://www.404media.co/hugging-face-is-hosting-5-000-nonconsensual-ai-models-of-real-people/

[59] 17 U.S. Code § 107

Some scholars observe that the use of copyrighted works to create GenAI differs from other uses where courts have found fair use, in part because GenAI, unlike indexing the web for search or identifying functional code to make games compatible with video game consoles, depends on the expressiveness of the works it is trained on. Whereas Google Search would work the same regardless of whether a website's content consisted of random words, GenAI only works well if it is trained on coherent, expressive, high-quality data. In other words, GenAI could not exist without exploiting the expressiveness of the works on which it is trained-[60]

## An Absurd Approach to Fair Use Analysis

My argument is a step removed from the one above. I posit that, whatever people may think about how the four factors apply to GenAI, any argument a GenAI company can make for fair use applies with equal or greater force to any individual human who wants to argue that their downloading, reading, watching, or listening to copyrighted works is fair use.[61] Human use is more transformative than GenAI use, for example. It takes far less data for humans to generalize and develop far more impactful insights, including all major scientific discoveries and genres of art. Additionally, humans are far less likely to engage in unfair competition with copyright owners because they are far less capable of memorizing and regurgitating copyrighted works. Therefore, if a court were to decide that a GenAI company's use is fair use, it must also support the notion that any human's use of copyrighted works without authorization is fair use as long as they do not produce an infringing output. This reductio ad absurdum reveals the weakness of the fair use argument for GenAI companies.

## The Scale of Copyright Infringement

Nearly all uses of copyrighted works by GenAI companies, whether open source or otherwise, should be considered infringement. Major GenAI datasets are composed of trillions of tokens scraped from the internet, a corpus that is overwhelmingly protected by copyright. The scale of this appropriation is extraordinary. For context, Ai2's OLMo 2 13B was trained on up to five trillion tokens,[62] Hugging Face's SmolLM2 was trained on approximately 11 trillion tokens,[63] and Meta's Llama 4 was trained on 30-60 trillion tokens.[64] These are just three of nearly two million models available on just a single website: Hugging Face.[65] Just the five most downloaded text-generation models (as in, not including image-to-text, image generation, text-to-image, and all other forms of GenAI) have been downloaded over 50 million times-

---

[60] There are other important distinctions. Google search connects creators with people they likely want to reach, for example, whereas GenAI often does not effectively connect people directly to creator content.

[61] David Atkinson, Unfair Learning: GenAI Exceptionalism and Copyright Law (unpublished 2025)

[62] OLMo 2 (7B and 13B), Allen Institute for Artificial Intelligence, *https://allenai.org/olmo*.

[63] Loubna Ben Allal et al., SmolLM2: When Smol Goes Big — Data-Centric Training of a Small Language Model, https://arxiv.org/html/2502.02737v1#:~:text=To%20attain%20strong%20performance%2C%20we,%2C%20and%20instruction%2Dfollowing%20data.

[64] *Kadrey v. Meta Platforms,* No. 3:23-cv-03417-VC-TSH, 2025 WL 56789 (N.D. Cal. Mar. 24, 2025), https://storage.courtlistener.com/recap/gov.uscourts.cand.415175/gov.uscourts.cand.415175.489.0.pdf

[65] Hugging Face, https://huggingface.co/models?sort=trending; Not all models on this site are GenAI models, but at least 100,000 are.

The sheer number of freely available models and their apparent popularity suggest that, in the aggregate, open-source models could cause at least as much harm to copyright owners as the proprietary models typically the focus of litigation. If courts require each version of open models to be litigated, the volume could quickly make litigation infeasible for all but the most well-endowed plaintiffs. Furthermore, some platforms, such as Amazon's Bedrock and Google's Vertex AI, make switching between models relatively easy for users, meaning that when one model is taken down for infringement, another could pop up to take its place, resembling a game of whack-a-mole, but perhaps more like whack-a-Hydra.

## Known Use of Pirated Content

Many of these datasets, used by both open and closed models, are known to contain material from notorious pirate sites. This is not inadvertent scraping but deliberate acquisition of content, mostly stolen from copyright owners, either by hacking the websites hosting the content, using stolen or fraudulently obtained credentials to log in without authorization, downloading the content, and moving it to the pirate website; or by accessing content with authorization, stripping it of its Digital Rights Management software, and then moving it to pirate sites.

From the litigation in *Kadrey v. Meta,* we know that Meta torrented hundreds of terabytes of content from pirate sites, including books and articles.[66] Even if it remains unclear whether open-source providers like Hugging Face or Ai2 have engaged in similar practices, the potential for infringement is evident because neither entity has unequivocally stated it doesn't and won't use such data. Their silence on this issue is deafening.

## Specific Examples of Problematic Datasets

A specific example of a problematic dataset is the Books3 dataset, supported by EleutherAI and comprising nearly 200,000 pirated books. Hugging Face hosted the Books3 link and, although aware of the dataset's contents, was reluctant to remove it. The website hosting the actual content ultimately removed the dataset link, not Hugging Face. This is concerning because it may amount to knowingly hosting a link to a dataset of pirated books, thereby facilitating actions that could constitute mass copyright infringement. As discussed below, Books3 is not the only incident of this kind on Hugging Face. This pattern of behavior is inconsistent with supporting the rights of copyright holders.

A similar concern arose with LAION-5B. It took years before researchers identified over a thousand instances of child sexual abuse material in a dataset that was widely used to train image generators[67] It demonstrated that the risks at issue are not limited to copyright but extend to privacy and bodily autonomy.

---

[66] Interestingly, of the over forty lawsuits against GenAI companies, only one defendant claims to be open source: Meta. It's unclear why other open-source providers have so far remained unscathed. In theory, it should be easier to prove infringement from reproduction (especially when the open-source entity releases the training dataset), derivatives (when then model regurgitates substantially similar outputs to the training data), and distribution (when copies are stored in the model, because the entity is giving those copies to others by releasing the models). For example, some researchers and law professors have shown that "Llama 3.1 70B memorizes some books, like Harry Potter and 1984, almost entirely."

[67] https://cyber.fsi.stanford.edu/news/investigation-finds-ai-image-generation-models-trained-child-abuse

## The "Publicly Available" Fallacy

GenAI companies often justify their practices by claiming the content is “publicly available.” This reasoning is deeply flawed. Almost all data is, or can be, made "publicly available” without the copyright owner’s permission. Public availability does not equate to placing material in the public domain; creators share their work to gain revenue, reputation, or exposure, not to forfeit their rights. Meta has taken this flawed reasoning to its logical extreme, arguing that even pirated data counts as “publicly available” because the pirate sites are accessible to anyone.[68] This reasoning would effectively nullify copyright protection for any work accessible online.

Similarly, when companies invoke “permissive licensing” for code, they overlook that permissive licensing is distinct from “unlicensed” or “without restrictions.” Even permissive licenses include binding legal obligations.

## Unique Challenges for Open-Source Models

Another copyright issue with open-source GenAI is that safeguards, such as filters, can be removed, making it impossible to prevent infringing outputs. For instance, blocking the ability to generate “in the style of” a particular artist won't work once the model is downloaded. This is a known issue.[69] This means that if the model can regurgitate infringing content, the open-source entities may be knowingly distributing copies of copyrighted works without an effective means to prevent infringement.

*Vicarious and Contributory Infringement*

Typically, for closed-source GenAI companies like OpenAI, the company would be required to take action to prevent future infringing outputs once made aware of them, or else it could be liable for vicarious or contributory copyright infringement. Vicarious infringement “allows for Party A to be found liable for the infringing acts of Party B if (i) Party A had the right and ability to control the infringing activity and (ii) Party A had a direct financial interest in the infringement.”[70] Contributory infringement “requires that (i) Party A makes a material contribution to the infringing activity, while (ii) having knowledge or a reason to know of the direct infringement by Party B.”

For open-source models, open-source entities cannot add patches or filters to downloaded models or datasets, nor can they delete them, even after being made aware that their model or dataset is infringing. However, the open-source company may still benefit financially, as Zuckerberg has noted that Meta benefits from releasing open-source artifacts in multiple earnings calls.[71] It cannot be the case that a company is no longer liable for infringing outputs simply because it has chosen to release a model and no longer controls it. The act of initial distribution with knowledge of likely infringement potential should establish liability.

---

[68] *Kadrey v. Meta Platforms,* No. 3:23-cv-03417-VC-TSH, 2025 WL 56789 (N.D. Cal. Mar. 28, 2025)., https://storage.courtlistener.com/recap/gov.uscourts.cand.415175/gov.uscourts.cand.415175.501.0.pdf (“To train Llama, Meta copied third party datasets containing Plaintiffs’ works from publicly available websites…”)

[69] *How to Regulate Unsecured AI*, CIGI (last visited July 7, 2025), https://www.cigionline.org/articles/not-open-and-shut-how-to-regulate-unsecured-ai/#:~:text=Ability%20to%20remove%20safety%20features,running%20on%20their%20own%20hardware.

[70] David Atkinson and Jacob Morrison, *A Legal Risk Taxonomy for Generative Artificial Intelligence*, (unpublished manuscript) https://arxiv.org/html/2404.09479v3

Some entities, like Hugging Face, don't merely release their own open-source artifacts; they also host open-source artifacts for others. As mentioned above, this meant they hosted the link to a dataset known to contain nearly 200,000 pirated books. They also host thousands of other datasets with mostly copyrighted material, including articles,[72] books,[73] webpages (like C4), lyrics,[74] scripts,[75] and code.[76] This is not inadvertent hosting—the information about the contents of these datasets is publicly available and well-documented.

The *Washington Post*, working in collaboration with Ai2 researchers, noted that the widely-used dataset called the Colossal Clean Crawled Corpus (C4) consisting of Common Crawl datasets (which, again, only collect samples from websites), which is one of the most popular training datasets and one hosted by Hugging Face, contains data scraped from at least 28 known notorious pirate sites.[77] In other words, Hugging Face is arguably facilitating copyright infringement by knowingly hosting such datasets. Moreover, they profit from doing so. They don't host the datasets out of the kindness of their hearts. Rather, they do it to generate revenue. As of August 2023, following their last funding round, they were valued at $4.5 billion. This substantial valuation demonstrates the commercial value derived, in part, from hosting potentially infringing content.

## Industry Acknowledgment of Copyright Concerns

While copyright seems like an obvious reason most companies don't share their training data, Ai2's CEO sees things differently. In response to the question "Why aren't more AI developers sharing training data for models they say are open?" in an interview with *Fast Company*, he replied:

"If I want to postulate, some of these training data have *a little bit of questionable material in them*. But also the training data for these models are the actual IP. The data is probably the most sacred part. Many think there's a lot of value in it. In my opinion, rightfully so. Data plays a significant role in improving your model,

---

71 Zuckerberg, M. & Dorell, K. Fourth Quarter 2023 Results conference call (2024); Zuckerberg, M. & Crawford, D. First Quarter 2023 Results conference call (2023) ("[PyTorch] has generally become the standard in the industry […] it's generally been very valuable for us […] Because it's integrated with our technology stack, when there are opportunities to make integrations with products, it's much easier to make sure that developers and other folks are compatible with the things that we need in the way that our systems work.")

72 *See, e.g.,* Andy Reas, *frontpage-news*, Hugging Face, https://huggingface.co/datasets/AndyReas/frontpage-news. (last visited July 7, 2025)

73 *See, e.g., Anti-Piracy Group Shuts Down Books3, a Popular Dataset for AI Models*, Interesting Eng'g (n.d.), https://interestingengineering.com/innovation/anti-piracy-group-shuts-down-books3-a-popular-dataset-for-ai-models [https://perma.cc/4P45-R4QJ].

74 *See, e.g.,* brunokreiner, *genius-lyrics*, Hugging Face, https://huggingface.co/datasets/brunokreiner/genius-lyrics.

75 *See, e.g.,* IsmaelMousa, *Movies*, Hugging Face, https://huggingface.co/datasets/IsmaelMousa/movies.

76 *See, e.g.*, BigCode, *starcoderdata*, Hugging Face, https://huggingface.co/datasets/bigcode/starcoderdata.

77 https://www.washingtonpopen-sourcet.com/technology/interactive/2023/ai-chatbot-learning/

changing the behavior of your model. It's tedious, it's challenging. Many companies spend a lot of dollars, a lot of investments, in that domain and they don't like to share it."[78] (emphasis added)

This candid admission reveals the industry's priorities: the copyrighted works of others are treated as less important than the trade secrets of the GenAI companies that exploit them. Copyright holders' rights matter very little to GenAI entities (even, apparently, in the estimation of an open-source entity leader). Yet, the companies go out of their way to protect their own intellectual property. This selective respect for intellectual property rights is both hypocritical and legally indefensible.

### The Takeaway

Where copyright law applies to closed-source models and datasets, it also applies to open-source models and datasets. Making GenAI artifacts open source does not meaningfully change the above analysis. Therefore, these entities should either refrain from using copyrighted content or obtain a license to use it.

## C. Terms of Service Violations

When scraping the web to create massive datasets, companies tend to scrape every page of the websites they visit. After all, all else being equal, more data is better than less data when training a model. Most sites have some terms of service or terms of use.[79] Some sites require users to accept these terms when they visit, thereby making them enforceable. This is the case when the user must click an "I Accept" box, for example, before accessing the site's content. This is known as "clickwrap." More often, sites include only a link to the terms at the bottom of the website. This is known as "browsewrap." As long as users don't click the link, the terms are typically unenforceable.

Clickwrap is a form of constructive notice. That is, courts presume you had a fair opportunity to review the terms before accepting them, thereby rendering them enforceable. You do not need constructive notice if there is actual notice. Actual notice occurs when you visit the terms rather than merely clicking "I accept" next to a hyperlink to the terms.

The issue is that when bots scrape every webpage (and even when they scrape only a sample of webpages, as Common Crawl does), including the terms pages, the bot deployer (i.e., the AI company or the company working on behalf of the AI company) then has actual notice of the terms, no different from when a human clicks on them. As described above, when you click the link or otherwise access the terms, they may become binding. This creates actual notice that should bind the scraping entity to the terms.

---

[78] Ali Farhadi, *AI2's Ali Farhadi Advocates for Open-Source AI Models. Here's Why*, Fast Company (Apr. 16, 2024), https://www.fastcompany.com/91283517/ai2s-ali-farhadi-advocates-for-open-source-ai-models-heres-why.

[79] Tangentially related, OpenAI has publicly stated that they are investigating and reviewing indications that DeepSeek, a Chinese AI startup, may have violated OpenAI's terms of service. The core of their claim revolves around a technique called "distillation," involves repeatedly querying OpenAI's models (like ChatGPT) at scale and using the generated outputs as training data for DeepSeek's own competing AI. OpenAI's terms of use explicitly prohibit users from "automatically or programmatically extract data or Output" and "use Output to develop models that compete with OpenAI." OpenAI believes DeepSeek's alleged distillation practices directly violate these contractual terms.

The prevalence of this issue is evident in the data itself. The C4 dataset contains 424 URLs from a "terms" page. This demonstrates that scraping bots routinely encounter and ingest terms of service pages, creating widespread constructive notice of usage restrictions.[80]

If the terms state, "You may not use any content on this website to train an AI system," and a GenAI company uses that content to train an AI system, it is likely in breach of the contract. Importantly, the law provides no exemption for open-source companies to ignore contract law. The open-source nature of the final product does not excuse the initial breach of contract.

## D. Licensing Violations and Attribution Failures

Code, datasets, and models are typically shared under licenses, which are contracts that outline how individuals and entities may use copyrighted material. Most licenses, including common open-source licenses like MIT and Apache 2.0,[81] are not a free-for-all; they come with conditions, most notably the requirement to provide attribution to the original creator. For example, here are the requirements under Apache 2.0:

**4. Redistribution**. You may reproduce and distribute copies of the Work or Derivative Works thereof in any medium, with or without modifications, and in Source or Object form, provided that You meet the following conditions:

- You must give any other recipients of the Work or Derivative Works a copy of this License; and
- You must cause any modified files to carry prominent notices stating that You changed the files; and
- You must retain, in the Source form of any Derivative Works that You distribute, all copyright, patent, trademark, and attribution notices from the Source form of the Work, excluding those notices that do not pertain to any part of the Derivative Works; and
- If the Work includes a "**NOTICE**" text file as part of its distribution, then any Derivative Works that You distribute must include a readable copy of the attribution notices contained within such NOTICE file, excluding those notices that do not pertain to any part of the Derivative Works, in at least one of the following places: within a NOTICE text file distributed as part of the Derivative Works; within the Source form or documentation, if provided along with the Derivative Works; or, within a display generated by the Derivative Works, if and wherever such third-party notices normally appear. The contents of the NOTICE file are for informational purposes only and do not modify the License. You may add Your own attribution notices within Derivative Works that You distribute, alongside or as an addendum to the NOTICE text from the Work, provided that such additional attribution notices cannot be construed as modifying the License.

---

[80] Moreover, C4 if just one of many datasets used to train models. No highly trained model is trained on just C4. In all likelihood, the additional datasets include additional terms webpages.

[81] According to the Open Source Initiative, "For Pypi, the package manager for Python [the most common language for AI research], components under the MIT and Apache 2.0 licenses dominate, at 29.14% and 23.98% respectively." Many open-source GenAI models are released under Apache 2.0., including those by Ai2, Mistral, LAION, Hugging Face, Salesforce, Amazon, and xAI.Ecosystem Graphs: The Social Footprint of Foundation Models, Stan. U. Inst. for Hum.-Centered AI (last visited July 30, 2025), https://crfm.stanford.edu/ecosystem-graphs/index.html?mode=table.

- You may add Your own copyright statement to Your modifications and may provide additional or different license terms and conditions for use, reproduction, or distribution of Your modifications, or for any such Derivative Works as a whole, provided Your use, reproduction, and distribution of the Work otherwise complies with the conditions stated in this License.

### The Attribution Problem in GenAI

Traditionally, adhering to these terms was trivial. If you used some code written by someone else, you could easily give them credit and use the appropriate license when you wrote your code. But training GenAI is a different beast. When a model generates code, it is incredibly difficult, if not impossible, to trace that output back to all its original licensed sources to provide the required attribution. This presents a problem for GenAI companies because the judge in *Doe 1 v. GitHub* allowed plaintiffs to pursue contract claims based on license violations, a claim for which fair use is not a defense.[82] In *Doe 1*, the plaintiffs, primarily software developers, allege that Copilot was trained on their copyrighted open-source code from public GitHub repositories without proper attribution or adherence to the terms of the open-source licenses (e.g., MIT, GPL, Apache). They claim that Copilot's generated code snippets are identical or substantially similar to their original work, constituting copyright infringement and a violation of license conditions for failing to provide attribution. The court's precedent holds that license violations can be pursued as contract claims, independent of copyright analysis.

Open-source licensing does not imply unrestricted use of code. Organizations such as EleutherAI or Ai2 may redistribute, modify, or create derivatives of open-source code, but they remain obligated to adhere to the original license terms. The popular Apache 2.0 license, for example, requires sharing a copy of the Apache 2.0 license, prominently notifying others of any changes to the files, and providing attribution to the code creators. Perhaps unsurprisingly, no popular open-source model complies with these open-source license terms. Instead, they argue that because the code was made public, they are permitted unrestricted use for training GenAI models. This argument misunderstands the nature of open-source licensing.

The inability to comply with a license should not be a legal excuse to disregard the creator's wishes and the license's requirements. Rather, it should mean that open-source GenAI companies must either find an alternative solution (e.g., paying to use the content without restrictions) or refrain from using it altogether. The technical difficulty of compliance does not excuse legal violations.

## E. Privacy Law Violations and Data Protection Failures

Three primary privacy issues with open-source GenAI could affect model inputs (the training data) and outputs (what the models generate): (i) current filtering techniques can reliably identify only a few types of personally identifiable information; (ii) widely releasing datasets disseminates any personally identifiable information (PII) they contain; and (iii) there is no way to remove PII from a downloaded open-source model once it has been released.

---

[82] *DOE 1 v. GitHub*, Inc., No. 4:22-cv-06823 (N.D. Cal. Feb. 11, 2025)

## Limited Effectiveness of PII Filtering

First, removing PII from training data is far more difficult than it sounds. A study by researchers from Ai2 and Hugging Face found that among eight common types of personal data (phone numbers, email addresses, US bank numbers, US social security numbers, IP addresses, credit card numbers, IBAN codes, and names), filtering tools could accurately and routinely identify only three: phone numbers, email addresses, and IP addresses.[83] This means that if a GenAI company wants to remove PII from its training data, it can do so with high accuracy for fewer than half of the most common types of personal information. Consequently, here is how Ai2 explained its process for filtering PII from its largest training dataset:

> Data sampled from the web can also leak personally identifiable information (PII) of users (Luccioni and Viviano, 2021; Subramani et al., 2023). Traces of PII are abundant in large-scale datasets (Elazar et al., 2023), and language models have also been shown to reproduce PII at inference time (Carlini et al., 2022; Chen et al., 2023b). Dolma's size makes it impractical to use model-based PII detectors like Presidio (Microsoft, 2018); instead, we rely on carefully-crafted regular expressions that sacrifice some accuracy for significant speed-up. Following Subramani et al. (2023), we focus on three kinds of PII that are detectable with high precision: email addresses, IP addresses and phone numbers. For documents with 5 or fewer PII spans, we replace the span with a special token (e.g., |||EMAIL_ADDRESS|||); this affects 0.02% of documents. Otherwise, we remove entire documents with higher density of PII spans; this affects 0.001% of documents. In data ablation experiments, we find that execution details around PII (e.g., removal versus special token replacement) had no effect on model performance, which is expected given the tiny percentage of affected data.[84]

## Amplified Risks in Open-Source Distribution

Second, for closed-source companies like OpenAI, the PII in their datasets remains hidden from public view, and they can at least attempt to filter PII from the model's output. With open source, these minimal safeguards all but disappear. Instead, anyone can access the datasets and view any PII therein. This can transform privacy violations from contained incidents into widespread data breaches, depending on which law(s) may apply. Because all safeguards can be easily and even unintentionally removed from models once they are downloaded, open-source entities cannot control what PII may be generated by the open-source models. Moreover, some entities, such as Hugging Face, host datasets comprising information scraped from social media sites, which are likely to contain personal data and enable others to infer additional information about users.[85]

---

[83] Nishant Subramani et al., *Detecting Personal Information in Training Corpora: an Analysis*, in Proc. of the 3d Workshop on Trustworthy Nat. Language Processing 2023, at 208-220 (Ass'n for Comput. Linguistics 2023), https://aclanthology.org/2023.trustnlp-1.18/.

[84] Luca Soldaini et al., Dolma: an Open Corpus of Three Trillion Tokens *for Language Model Pretraining Research*, in Proc. of the 62nd Ann. Meeting of the Ass'n for Comput. Linguistics (Volume 1: Long Papers) 15725, 15725–15788 (Ass'n for Comput. Linguistics 2024), https://aclanthology.org/2024.acl-long.840/.

### The Impossibility of Data Deletion

Third, removal and suppression techniques for GenAI models are severely limited.[86] It is challenging or impossible to determine where a specific type of PII for an individual resides within massive GenAI models, and it can also be difficult or impossible to discern the relationship between model weights and model outputs. This makes compliance with data deletion requirements all but impossible. Once personal information is embedded in a model's weights and that model is distributed, the data essentially becomes unretractable.

This leaves a stark choice: either GenAI companies must risk knowingly violating privacy laws, or society must weaken its privacy laws to accommodate AI's technical limitations. The question is whether we value this specific form of technological progress more than the right to privacy.

## F. Law Conclusion

The legal analysis shows that open-source GenAI faces many of the same risks and responsibilities as proprietary systems, and that exemptions based solely on openness are both problematic and potentially harmful. Beyond the law, however, open-source advocates often invoke broader public policy arguments to justify special treatment. The following section evaluates these policy claims, questioning whether open-source GenAI truly delivers on its promises of market disruption, democratization, and societal benefit.

# VI. Public Policy and Open Source

Beyond purely legal arguments, open-source GenAI entities often advance policy claims to justify exemptions from existing laws and regulations. Even when their practices do not align with the letter and spirit of applicable law, proponents argue that their work serves vital societal purposes. These advocates contend that open-source practitioners serve the greater good, even when their methods may undermine the intellectual property, contractual, and privacy rights of content creators, thereby undermining the very system that incentivizes the creation and sharing of the data these GenAI systems require for training. For such proponents, the perceived benefits are so blindingly obvious and certain that ignoring the law is deemed entirely justified.

Open-source entities also strive to match the capabilities of proprietary, state-of-the-art models. This mimicry is problematic because GenAI entities rely on model evaluations to demonstrate progress and compare models. However, these evaluations tell us little about the ethical or responsible development of models. Instead, model evaluation tends to focus on qualities most likely to influence adoption, such as toxicity, accuracy, bias, and user-friendliness. Few people will choose one model over another solely because it produces 10% fewer carbon emissions, but most users will opt for a model that is 10% more accurate. This adoption preference explains why the benchmarks most revered by GenAI entities are MMLU, HellaSwag, GSM8K, and ARC-C, rather than results from third-party environmental impact audits or assessments of the amount of publicly available, properly licensed data used for training. This emphasis on adoption may also explain why open-source GenAI organizations frequently lack public ethics charters and, even more rarely, enforceable ethical commitments.

---

[85] Samantha Cole, *Someone Made a Dataset of One Million Bluesky Posts for Machine Learning Research*, 404 Media (Oct. 18, 2023), https://www.404media.co/someone-made-a-dataset-of-one-million-bluesky-posts-for-machine-learning-research/.

[86] See, e.g., A. Feder Cooper et al, Machine Unlearning Doesn’t Do What You Think:
*Lessons for Generative AI Policy, Research, and Practice*, Preprint. Prior version presented at the 2nd Workshop on Generative AI + Law at ICML ’24, https://arxiv.org/pdf/2412.06966

Furthermore, the argument that open-source GenAI must be built irresponsibly, unethically, and illegally because it is the only way to understand state-of-the-art LLMs is analogous to a biologist claiming they shouldn't need to experiment on worms or mice, only on humans, if they are ever to understand important aspects of human biology. Instead, the open-source community could make the most of the vast amount of public-domain and properly licensed material. The practice of risking economic and environmental damage should not be taken lightly or assumed to be a given when conducting important and fundamental research.

Arguments frequently advanced by open-source proponents center on marketplace advantages, the democratization of a technology likened to humanity's most profound discoveries, environmental benefits, an unquestioning belief in technological progress, and the claim that open source makes GenAI safer. This paper will critically examine each of these claims, demonstrating why such policy arguments do not warrant special legal exemptions for open-source GenAI. To be clear, the point of this section is not to say that there are no benefits from closed-source or open-source AI, or that open-source AI is categorically bad. Rather, the point is that there are no overwhelming benefits from open-source AI that justify relaxed ethical and legal standards.

First, this analysis will address claims that open-source GenAI inherently leads to market disruption and democratization. It will examine whether these assertions hold in absolute terms. If open-source models do not demonstrably increase market competition or democratize GenAI, then these arguments cannot justify preferential treatment.

Next, the article will pivot to a comparative analysis, evaluating whether open-source GenAI is truly superior to its proprietary counterparts in terms of environmental impact, innovation, and safety. This section will operate in relative terms because a lack of empirical data makes a review based on absolute terms impossible, and advocates for legal exemptions for open-source models implicitly argue for the inherent superiority of open-source models over closed-source alternatives. Therefore, for these policy arguments to justify such exemptions, they must convincingly demonstrate that open-source GenAI offers a marked advantage over proprietary options in these areas.

## A. Oligopolies and Open-Source GenAI

A prevalent but flawed assumption holds that open-source GenAI will necessarily disrupt entrenched market power and foster meaningful competition. In practice, the infrastructure, data, and capital required to develop and deploy state-of-the-art GenAI remain concentrated among a handful of dominant technology firms, limiting the competitive impact of open-source initiatives.

Illustrating this perspective, Wired profiled Shawn Presser, the creator of the Books3 dataset at the center of some of the most consequential ongoing litigation. According to the article:

> In his eyes, it leveled the playing field for smaller companies, researchers, and ordinary people who wanted to create large language models. He believes people who want to delete Books3 are unintentionally advocating for a generative AI landscape dominated solely by Big Tech-affiliated companies like OpenAI. "If you really want to knock Books3 offline, fine. Just go into it with eyes wide open. The world that you're choosing is one where only billion-dollar corporations are able to create these large language models," he says.[87]

While the idea of leveraging open source to disrupt established market power sounds compelling and noble in theory, it fails in practice. Books3 became available in October 2020, yet nearly five years later, only a handful of highly capitalized companies, including OpenAI, Google, Meta, and Microsoft, continue to dominate the GenAI landscape.

When commentators discuss open source as a means to disrupt large tech companies, they often envision it as the triumph of a scrappy underdog. However, history is replete with examples of for-profit companies using open-source software to gain or maintain their competitive edge. In other words, being open-source, per se, is not the primary driver of market disruption. Instead, it is a large, mostly closed-source for-profit company using open-source as a tool to take on another large, mostly closed-source for-profit.

For example, numerous for-profit entities have funded open-source projects to counter their competitors. Consider IBM's billion-dollar strategic investment in Linux. It was a move not driven by pure altruism but by a desire to challenge its chief business adversary, Microsoft. An even more striking example is Google's Android operating system: by investing heavily in Android, Google transformed it into the world's most widely used mobile operating system, with each iteration reinforcing its ecosystem in its competition with Apple. Google Chrome's development follows a similar strategic narrative in its competition with Internet Explorer.

This pattern is the norm rather than the exception when Big Tech is involved. Across multiple key technology sectors, despite the disruptive potential of open-source innovation, market concentration remains a dominant force. Consider these examples:

| Sector | Dominant Players | Market Share/Notes |
| --- | --- | --- |

[87] Emma Barnett, *The Battle Over Books is Being Fought on Three Fronts*, Wired (July 7, 2025), https://www.wired.com/story/battle-over-books3/.

| Cloud Services | AWS (Amazon), Azure (Microsoft), Google Cloud (Google) | ~63% market share[88] |
|---|---|---|
| Mobile Operating Systems | Android (Google), iOS (Apple) | ~99.5% market share[89] |
| Desktop Computers Operating Systems | Windows (Microsoft), iOS (Apple) | ~86% market share[90] |
| Datacenter AI Chips | Nvidia, Intel, AMD | ~65%, 22%, and 11% respectively[91] |
| GenAI Platforms | ChatGPT (OpenAI), Microsoft Copilot, Gemini (Google) | ~87% market share[92] |
| Social Media | Facebook/Instagram (Meta), TikTok (ByteDance), YouTube (Google) | Concentrated among a few large firms |

Another striking example of open source benefiting tech giants is AI development frameworks. These frameworks offer significant strategic advantages to their corporate creators, such as Meta (PyTorch) and Google (TensorFlow), by accelerating AI development and ensuring robust, predictable deployment. These companies leverage their frameworks to standardize AI development, creating modular, "Lego-like" systems that integrate seamlessly with their proprietary platforms. Google's TensorFlow exemplifies this strategy by optimizing for its Tensor Processing Unit (TPU) hardware, thereby bolstering its dominance in cloud AI computing. Moreover, these frameworks serve as gateways to profitable services, allowing companies to shape researchers' and developers' workflows and, ultimately, define the trajectory of the AI field by cultivating user familiarity with their preferred ecosystem and seamlessly integrating externally developed innovations into their proprietary, commercially driven products. This approach effectively outsources innovation and internal research and development.[93]

[88] *Worldwide market share of leading cloud infrastructure service providers*, Statista (2019), https://www.statista.com/chart/18819/worldwide-market-share-of-leading-cloud-infrastructure-service-providers/.
[89] *Mobile Operating System Market Share Worldwide*, StatCounter Global Stats (Feb. 2025), https://gs.statcounter.com/os-market-share/mobile/worldwide/#monthly-202502-202502-bar.
[90] *Desktop Operating System Market Share Worldwide*, StatCounter Global Stats (Feb. 2025), https://gs.statcounter.com/os-market-share/desktop/worldwide/#monthly-202502-202502-bar.
[91] Data-Center AI Chip Market – Q1 2024 Update, TechInsights (last visited July 7, 2025), https://www.techinsights.com/blog/data-center-ai-chip-market-q1-2024-update.
[92] Evan Bailyn, *Top Generative AI Chatbots by Market Share*, First Page Sage (Sept. 2024), https://firstpagesage.com/reports/top-generative-ai-chatbots/.

There is no compelling reason to believe that the trend of oligopolies dominating every profitable space will change because of the emergence of open-source GenAI projects. Although open source offers genuine opportunities for customization, most consumers and small businesses prioritize convenience and reliability. In practice, using a production-ready API, such as those provided by OpenAI, is typically more practical and cost-effective than developing or maintaining an open-source GenAI model independently. Thus, market forces continue to favor well-resourced, for-profit enterprises.

Moreover, Big Tech wields significant soft power. By strategically funding select researchers, sponsoring influential conferences, and shaping academic discourse, companies such as Google, Microsoft, Nvidia, and Amazon effectively determine which research questions are pursued and which criticisms are voiced. As GenAI research becomes increasingly expensive and researchers become more dependent on Big Tech funding, these advantages further consolidate Big Tech's market dominance.

The following table shows the valuations of major players in the GenAI space before and after the debut of ChatGPT, which sparked the chatbot arms race. Predictably, the tech giants' valuations ballooned, far outpacing the gains of GenAI startups for reasons detailed below[94]

| Company | Products | Valuation Q3 2022 | Valuation 11/30/2025 |
|---|---|---|---|
| Tech Giants | | | |
| Microsoft | CoPilot, Phi, Azure | $1.7 trillion[95] | $3.66 trillion |

[93] Zuckerberg, M. & Crawford, D. First Quarter 2023 Results conference call (2023) ("[PyTorch] has generally become the standard in the industry […] it's generally been very valuable for us […] Because it's integrated with our technology stack, when there are opportunities to make integrations with products, it's much easier to make sure that developers and other folks are compatible with the things that we need in the way that our systems work."), https://s21.q4cdn.com/399680738/files/doc_financials/2023/q1/META-Q1-2023-Earnings-Call-Transcript.pdf

[94] It's true that the relative rate of growth is faster for the startups, but this is because it's easier to go from a small number to a bigger one than from a huge number to a huger one. If you have a baby and it grows from five pounds to ten pounds, that's a 200% growth, even though it only added five pounds! If you weigh 200 pounds and you add fifty more, that's just a 25% increase. But another way to think about it is: would you rather have an extra $300 billion or an extra $2 trillion? Finally, it's notable that, except for Hugging Face, only the Big Tech companies are profitable.

[95] *Microsoft Market Cap,* Macrotrends, https://www.macrotrends.net/stocks/charts/MSFT/microsoft/market-cap (last visited July 7, 2025).

| amazon | Titan, Bedrock | $1.1 trillion[96] | $2.49 trillion |
|---|---|---|---|
| Google | Gemini, Google Cloud | $1.2 trillion[97] | $3.86 trillion |
| Meta | Llama | $363 billion | $1.63 trillion |
| GenAI Startups | | | |
| OpenAI | ChatGPT, GPT-4, DALL-E | $1.5 billion[98] | $500 billion |
| | Grok | Did not exist | $230 billion (targeted) |
| ANTHROP\C | Claude | $5 billion[99] | $230 billion (targeted) |
| MISTRAL AI_ | Mistral Large, Mixtral | Did not exist. | $14 billion |

[96] *Amazon Market Cap*, Macrotrends (last visited July 7, 2025), https://www.macrotrends.net/stocks/charts/AMZN/amazon/market-cap.

[97] *Google Market Cap*, Macrotrends (last visited July 7, 2025), https://www.macrotrends.net/stocks/charts/GOOGL/alphabet/market-cap

[98] Joanna Glasner, *The Biggest Funding Rounds for OpenAI and Silicon Ranch This Week*, Crunchbase News (Mar. 1, 2023, 11:24 AM PST), https://news.crunchbase.com/venture/biggest-funding-rounds-openai-silicon-ranch/#:~:text=If%20that%20occurs%2C%20OpenAI%20will%20have%20a%20post%2Dmoney%20valuation%20of%20%24300%20billion.&text=The%20company%20raised%20a%20%24103%20million%20Series,has%20raised%20%24350%20million%2C%20per%20the%20company.

[99] *Google Invests $300 Million in Anthropic as Race to Compete with ChatGPT Heats Up*, VentureBeat (last visited July 7, 2025), https://venturebeat.com/ai/google-invests-300-million-in-anthropic-as-race-to-compete-with-chatgpt-heats-up/.

| cohere | Command R | Unknown (less than $2 billion)[100] | $6.8 billion |
|---|---|---|---|
| Hugging Face | SmolLM | $2.2 billion[101] | $4.5 billion[102] |

This data suggests that the CEO of Ai2 may have been overzealous about the headway open source can make. When DeepSeek's "reasoning" model debuted, he was quick to proclaim that "We're celebrating this moment as a proof point that open source will win."[103] However, the evidence suggests that the Big Tech oligopoly is actually winning.[104] But even among small tech companies, the money mostly flows to just a handful of companies. As Axios notes, "In Q2, more than one-third of all U.S. venture dollars went to just five companies."[105]

The enduring market dominance of well-resourced companies stems from their role as indispensable suppliers of essential components in the GenAI supply chain. Multiple levers favor Big Tech in the GenAI market, and none appears likely to be significantly challenged by open-source initiatives:

| Lever | Notes |
|---|---|
| Network effects | Occurs when the value of a product or service increases as more people use it. Essentially, each new user adds value for existing users, creating a positive feedback loop that benefits all parties. This is the case with Hugging Face, where more users hosting and sharing more AI artifacts make the platform more valuable and useful. |
| Access to proprietary datasets | Companies such as Google (via YouTube), xAI (via X), and Meta (via Facebook and Instagram) benefit from exclusive datasets that power their AI models. |

---

[100] Cohere, *Wikipedia*, https://en.wikipedia.org/wiki/Cohere (last visited July 7, 2025).

[101] TechCrunch, Aug. 24, 2023, https://techcrunch.com/2023/08/24/hugging-face-raises-235m-from-investors-including-salesforce-and-nvidia/.

[102] Kyle Wiggers, *Hugging Face raises $235M from investors, including Salesforce and Nvidia*, TechCrunch (Aug. 24, 2023, 7:00 AM PDT), https://techcrunch.com/2023/08/24/hugging-face-raises-235m-from-investors-including-salesforce-and-nvidia/.

[103] Ali Farhadi, *Open Source Will Win: Allen Institute for AI CEO Ali Farhadi on the New Era of Artificial Intelligence*, GeekWire (Feb. 20, 2025), https://www.geekwire.com/2025/open-source-will-win-allen-institute-for-ai-ceo-ali-farhadi-on-the-new-era-of-artificial-intelligence/ (last visited July 7, 2025).

[104] Nvidia's stock dipped when DeepSeek debuted its models, but the stock recovered and is now at an all-time high, again demonstrating how undisruptable the tech giants are.

[105] Dan Patrick, *AI is Eating Venture Capital, or at Least its Dollars*, Axios (Jul. 3, 2025), https://www.axios.com/2025/07/03/ai-startups-vc-investments

| | |
|---|---|
| Capital resources | Training cutting-edge models, such as Gemini Ultra 1.0, which was estimated to cost $191 million to train[106], is feasible only for companies with deep pockets. Moreover, many GenAI startups are not yet profitable.[107] |
| Access to compute | Firms like xAI claim access to over 200,000 GPUs,[108] while Meta asserts that its infrastructure rivals the capacity of 600,000 H100 GPUs.[109] |
| No known business model can usurp power from Big Tech | There is no known business model capable of eroding Big Tech's integrated ecosystems; for example, it seems unlikely that Perplexity could unseat Google in the search market. |
| High-interest investment environment | This makes it more expensive for smaller companies to get the capital they need in order to develop GenAI that can compete with Big Tech |
| Access to markets: Google, Apple, Amazon, Meta, X, Samsung, etc. | Big Tech companies routinely secure preferential agreements, such as early access or discounted pricing, to integrate external models into their ecosystems. Examples include:<br>● OpenAI and Snapchat[110]<br>● xAI and Telegram[111]<br>● Gemini and Samsung (though Samsung may swap in Perplexity)[112] |

[106] Stanford HAI, *2025 AI Index Report*, https://hai.stanford.edu/ai-index/2025-ai-index-report.

[107] https://techcrunch.com/2025/01/05/openai-is-losing-money-on-its-pricey-chatgpt-pro-plan-ceo-sam-altman-says/; https://venturebeat.com/ai/anthropic-raises-3-5-billion-reaching-61-5-billion-valuation-as-ai-investment-frenzy-continues/#:~:text=Revenue%20skyrockets%201%2C000%25%20year%2Dover,revenue%20multiple%20shows%20market%20maturation

[108] Lora Kolodny, *Elon Musk: xAI and Tesla to keep buying Nvidia, AMD chips*, CNBC (May 20, 2025, 4:28 PM EDT), https://www.cnbc.com/2025/05/20/elon-musk-says-he-expects-to-keep-buying-gpus-from-nvidia-and-amd.html#:~:text=If%20that%20occurs%2C%20OpenAI%20will%20have%20a%20post%2Dmoney%20valuation%20of%20$300%20billion.&text=The%20company%20raised%20a%20$103%20million%20Series,has%20raised%20$350%20million%2C%20per%20the%20company.

[109] https://engineering.fb.com/2024/03/12/data-center-engineering/building-metas-genai-infrastructure/#:~:text=The%20future%20of%20Meta's%20AI%20infrastructure&text=future%20of%20AI.-,By%20the%20end%20of%202024%2C%20we're%20aiming%20to%20continue,equivalent%20to%20nearly%20600%2C000%20H100s.

[110] Aisha Malik, *Snapchat Launches an AI Chatbot Powered by OpenAI's GPT Technology*, TechCrunch (Feb. 27, 2023), https://techcrunch.com/2023/02/27/snapchat-launches-an-ai-chatbot-powered-by-openais-gpt-technology/.

[111] Rita Liao, *xAI to pay Telegram $300M to integrate Grok into app*, TECHCRUNCH (May 28, 2025, 1:04 PM PDT), https://techcrunch.com/2025/05/28/xai-to-pay-300m-in-telegram-integrate-grok-into-app/.

[112] Wes Davis, *Google is paying Samsung an 'enormous sum' to preinstall Gemini*, The Verge (Apr 26, 2025), https://www.theverge.com/news/652746/google-samsung-gemini-default-placement-antitrust-trial (last visited July 7,

| | |
|---|---|
| | ● Perplexity and Motorola[113] and PayPal[114] |
| Access to top talent | The advanced expertise required to develop leading GenAI systems is in scant supply. In response, companies like Meta have offered compensation packages ranging from $10 million to $100 million. Startups cannot match these amounts.[115] |

Ultimately, the companies best positioned to dominate GenAI are those with control over the foundational infrastructure required for training, deployment, and commercialization. This reality becomes clear when examining how just a handful of the largest companies continue to dominate key segments of the GenAI development supply chain:

| | |
|---|---|
| Compute Providers | ● Amazon (AWS)<br>● Google (Google Cloud)<br>● Microsoft (Azure) |
| Proprietary Platforms for Data | ● Amazon (Amazon.com)<br>● Google (YouTube, Google search index)<br>● Meta (Facebook, Instagram, Threads)<br>● Microsoft (Bing search index) |
| Model Development | ● Amazon (Titan)<br>● Apple (MM1)<br>● Google (Gemini, Gemma)<br>● Meta (Llama)<br>● Microsoft (Phi) |

---

2025).; Jay McGregor, *Samsung Galaxy Phone, Tablet Free Perplexity AI Offer*, Forbes (June 23, 2025), https://www.forbes.com/sites/jaymcgregor/2025/06/23/samsung-galaxy-phone-tablet-free-perplexity-ai-offer/.

[113] Perplexity AI, *Announcing Our Global Partnership with Motorola*, perplexity.ai (last visited July 7, 2025), https://www.perplexity.ai/hub/blog/announcing-our-global-partnership-with-motorola#:~:text=Announcing%20Our%20Global%20Partnership%20with%20Motorola&text=We're%20excited%20to%20announce,our%20answer%20engine%20and%20assistant.

[114] *Perplexity and PayPal Partner to Provide Easy Payment Checkouts for Users*, REUTERS (May 14, 2025), https://www.reuters.com/business/media-telecom/perplexity-paypal-partner-provide-easy-payment-checkouts-users-2025-05-14/.

[115] Cade Metz and Mike Isaac, *Meta Debuts AI Lab to Pursue Superintelligence*, N.Y. Times (June 10, 2025), https://www.nytimes.com/2025/06/10/technology/meta-new-ai-lab-superintelligence.html.; Erin Woo, *Inside the Great AI Talent Auction: The Deals, the Free Agents—and the Egos*, The Information (June 27, 2025, 9:04 AM PDT), https://www.theinformation.com/articles/inside-great-ai-talent-auction-deals-free-agents-egos?utm_source=ti_app.

| GenAI Partnerships / Investments / Agreements | ● Amazon (Anthropic[116], Mistral AI[117], Cohere[118], Hugging Face[119])<br>● Google (Anthropic[120], Mistral AI[121], Cohere[122], Ai2[123], Hugging Face[124])<br>● Microsoft (OpenAI[125], Mistral AI[126], xAI,[127] Cohere[128]) |
|---|---|
| Model Deployment Platforms | ● Amazon (Bedrock)<br>● Google (Vertex AI)<br>● Microsoft (Azure ML Studio) |

[116] Amazon, *Amazon completes $4B Anthropic investment to advance generative AI*, About Amazon (Mar. 27, 2024), https://www.aboutamazon.com/news/company-news/amazon-anthropic-ai-investment.

[117] *Amazon Bedrock adds Mistral AI models*, About Amazon (Mar. 1, 2024), https://www.aboutamazon.com/news/aws/mistral-ai-amazon-bedrock.

[118] Cohere, *Cohere Brings its Enterprise AI Offering to Amazon Bedrock*, Cohere (July 26, 2023), https://txt.cohere.com/amazon-bedrock/.

[119] Amazon, *AWS and Hugging Face collaborate to make generative AI more accessible and cost-efficient*, Aws Machine Learning Blog (Feb. 21, 2023), https://aws.amazon.com/blogs/machine-learning/aws-and-hugging-face-collaborate-to-make-generative-ai-more-accessible-and-cost-efficient/.

[120] Google Cloud, *Google Announces Expansion of AI Partnership with Anthropic*, Google Cloud Press Corner (Nov. 8, 2023), https://www.googlecloudpresscorner.com/2023-11-08-Google-Announces-Expansion-of-AI-Partnership-with-Anthropic.

[121] PYMNTS, Mistral AI Partners With Google Cloud to Distribute AI Solutions, Pymnts.com (Dec. 13, 2023), https://www.pymnts.com/news/artificial-intelligence/2023/mistral-ai-partners-with-google-cloud-distribute-solutions/

[122] Cohere, *Cohere is Available on the Google Cloud Marketplace*, TXT (last visited July 7, 2025), https://txt.cohere.com/cohere-is-available-on-the-google-cloud-marketplace/.

[123] Ai2, *Allen Institute for AI Announces Partnership with Google Cloud to Accelerate Open AI Innovation*, BusinessWire (Apr. 8, 2025), https://www.businesswire.com/news/home/20250408026138/en/Ai2-Allen-Institute-for-AI-Announces-Partnership-with-Google-Cloud-to-Accelerate-Open-AI-Innovation.

[124] Google Cloud & Hugging Face, *Google Cloud and Hugging Face Announce Strategic Partnership to Accelerate Generative AI and ML Development*, Pr Newswire (Jan. 25, 2024), https://www.prnewswire.com/news-releases/google-cloud-and-hugging-face-announce-strategic-partnership-to-accelerate-generative-ai-and-ml-development-302044380.html.

[125] Microsoft, *Microsoft and OpenAI Extend Partnership*, Microsoft On the Issues (Jan. 23, 2023), https://blogs.microsoft.com/blog/2023/01/23/microsoftandopenaiextendpartnership/.

[126] Microsoft, *Microsoft and Mistral AI announce new partnership to accelerate AI innovation and introduce Mistral Large first on Azure*, Azure Blog (Feb. 26, 2024), https://azure.microsoft.com/en-us/blog/microsoft-and-mistral-ai-announce-new-partnership-to-accelerate-ai-innovation-and-introduce-mistral-large-first-on-azure/.

[127] Erin Woo, *Microsoft Prepares to Host xAI's Grok Model on Azure*, The Information (May 1, 2025), https://www.theinformation.com/briefings/microsoft-prepares-host-xais-grok-model-azure.

[128] Jessica Toonkel & Aaron Holmes, *Microsoft to Sell AI From Cohere as Business Moves Beyond OpenAI*, The Information (Nov. 16, 2023), https://www.theinformation.com/briefings/microsoft-to-sell-ai-from-cohere-as-business-moves-beyond-openai

| | |
|---|---|
| Self-Serving Market Access | ● Google Gemini in Google Workspace[129]<br>● Microsoft Copilot in Microsoft 365[130] and Windows 11[131] |

Notably, research-focused organizations such as Ai2, Stanford, and similar academic institutions lack sufficient quantities of all the key elements listed above to successfully compete with tech giants. The same limitation likely applies to virtually everyone who downloads artifacts created and released by open-source entities. In short, radical openness is unlikely to meaningfully affect the existing power structure. The appealing rhetoric heard at high-level industry and thought leadership conferences and fireside chats cannot overcome the market forces that sustain the current technology ecosystem.

As noted in "Why Open AI Systems are Actually Closed and Why This Matters,"

> [p]inning our hopes on 'open' AI in isolation will not lead us to that world, and in many respects could make things worse, as policymakers and the public put their hope and momentum behind open AI, assuming that it will deliver benefits that it cannot offer in the context of concentrated corporate power.[132]

[129] Google Workspace, *Introducing Gemini for Google Workspace*, Google Workspace Blog (Feb. 21, 2024), https://workspace.google.com/blog/product-announcements/gemini-for-google-workspace.
[130] Microsoft, *Introducing Microsoft 365 Copilot – your copilot for work*, The Official Microsoft Blog (Mar. 16, 2023), https://blogs.microsoft.com/blog/2023/03/16/introducing-microsoft-365-copilot-your-copilot-for-work/.
[131] Microsoft, *Microsoft Copilot improvements for Windows 11*, Windows Experience Blog (Feb. 29, 2024), https://blogs.windows.com/windowsexperience/2024/02/29/microsoft-copilot-improvements-for-windows-11/.
[132] Jeffrey I. Katz et al., *Computing the Future: The need for sustainable and equitable access to large-scale AI infrastructure*, 628 Nature 238 (2024), https://www.nature.com/articles/s41586-024-08141-1.

## B. Democratization and Open-Source GenAI

Numerous entities claim to be democratizing GenAI in some fashion, including Meta,[133] the National Artificial Intelligence Research Resource (NAIRR),[134] Amazon,[135] H2O.ai,[136] Microsoft,[137] Hugging Face,[138] Stability AI,[139] and Google.[140] Yet, these assertions often amount to rebranding select open-source initiatives rather than achieving genuine democratization.[141] While openness is a key component, it represents only a means to an end—a necessary but insufficient condition for genuine democratization.

Claims of democratizing GenAI through open-source releases often overlook critical barriers to meaningful access. Merely making code or models available does not democratize technology if the vast majority of people lack the expertise, resources, or infrastructure to use or adapt these tools. True democratization requires not only access but also education, support, and participatory decision-making. Instead, most acts of "democratizing AI" focus on democratizing AI for people who already possess relevant expertise or have the time and resources to acquire the necessary skills. This population is probably fewer than 100,000 people globally, or a fraction of a percent of the world's population.[142]

---

[133] Meta, *Democratizing Access to Large-Scale Language Models with OPT-175B*, Meta AI, https://ai.meta.com/blog/democratizing-access-to-large-scale-language-models-with-opt-175b/ (last visited July 7, 2025); Meta, *Democratizing Conversational AI Systems Through New Data Sets and Research*, Meta AI, https://ai.meta.com/blog/democratizing-conversational-ai-systems-through-new-data-sets-and-research/ (last visited July 7, 2025).

[134] CIO, Forbes Staff and Megan Poinski, *New Federal Program Aims To Democratize AI Research And Development*, Forbes (Jan. 25, 2024), https://www.forbes.com/sites/cio/2024/01/25/new-federal-program-aims-to-democratize-ai-research-and-development/; Will Henshall, *The U.S. Just Made a Crucial Step Toward Democratizing AI Access*, TIME (last visited July 7, 2025), https://time.com/6589134/nairr-ai-resource-access/.

[135] Amazon, *Siemens and AWS Join Forces to Democratize Generative AI in Software Development*, aboutamazon.com, (Mar. 2024), https://press.aboutamazon.com/aws/2024/3/siemens-and-aws-join-forces-to-democratize-generative-ai-in-software-development; Amazon, *Democratizing generative AI to inspire diverse innovation: Insights from AWS executive Dave Levy*, AWS Public Sector Blog, aws.amazon.com, https://aws.amazon.com/blogs/publicsector/democratizing-generative-ai-to-inspire-diverse-innovation-insights-from-aws-executive-dave-levy/

[136] H2O.ai, *Democratizing AI*, H2O.ai, https://h2o.ai/insights/democratizing-ai/ (last visited July 7, 2025).

[137] Microsoft, *Democratizing AI - Stories*, *Microsoft News* (last visited July 7, 2025), https://news.microsoft.com/features/democratizing-ai/; Microsoft, *Democratizing AI: Satya Nadella on AI Vision and Societal Impact at DLD*, *Microsoft News* (Jan. 17, 2017), https://news.microsoft.com/europe/2017/01/17/democratizing-ai-satya-nadella-shares-vision-at-dld/.

[138] Hugging Face, *Intel and Hugging Face Partner to Democratize Machine Learning Hardware Acceleration*, Hugging Face (Mar. 28, 2023), https://huggingface.co/blog/intel.

[139] Emad Mostaque, *Democratizing AI, Stable Diffusion & Generative Models*, Scale Events (2022), https://exchange.scale.com/public/videos/emad-mostaque-stability-ai-stable-diffusion-open-source.

[140] Google, *Democratizing Access to AI-Enabled Coding with Colab*, Google AI Blog (Mar. 23, 2023), https://blog.google/technology/ai/democratizing-access-to-ai-enabled-coding-with-colab/.

[141] Merriam-Webster defines democratization as "to make democratic." The version of "democratic" at issue here is "relating, appealing, or available to the broad masses of the people: designed for or liked by most people" and "especially: organized or operated so that all people involved have power, influence, etc."

[142] Gradient Flow, *Where Do Machine Learning Engineers Work?*, https://gradientflow.com/where-do-machine-learning-engineers-

In other words, GenAI is democratized only if a broad enough segment of the public can use it effectively. Simply making GenAI artifacts available to those without the requisite technical knowledge does not meaningfully democratize access. True democratization requires that a broad segment of the public can effectively use, adapt, and understand the technology. Furthermore, if the model is trained exclusively on English text (or on a small number of languages), so that its outputs are primarily in English, then the model serves only those literate in English, which hardly seems democratic when considering a global audience.[143] For context, only approximately 17% of the world's population *speaks* English, and not all speakers can *read* it.[144]

Data from Cloudflare further illustrates this disparity: only one country accounted for more than 10% of internet traffic to GenAI websites in March 2025 — the United States, at 22.7%. The next closest country accounted for only 8.3%. Of the countries representing at least 3% of the traffic, four were newly industrialized nations, and five were fully industrialized. The first African country doesn't appear on the list until number 32–Egypt–with a 0.7% share.[145]

GenAI is not democratized if it remains inscrutable to most people and if they lack a voice in what is being built or how data is used. True democratization of GenAI demands more than mere access or basic literacy. It requires comprehensive education about how GenAI works, its applications, and its societal impact. Democratization should empower people not only to interact with basic products like chatbots but also to download, modify, and even fine-tune GenAI models. It must also enable non-technologists to understand what's happening, which data are being used, and the rationale behind methodological choices.[146] The mere fact that thousands of open-source models exist on platforms like Hugging Face means little to almost everyone globally. Similarly, a technical paper intended solely for experts contributes little to broader knowledge dissemination; rather, information must be communicated in accessible terminology, with relatable examples and clear analogies that resonate with a broader audience.[147]

---

work/#:~:text=There%20are%20over%2010%2C000%20people,their%20current%20or%20previous%20roles (last visited July 7, 2025).); the conservative 100,000 number used in this paper is 5x what Gradient Flow indicates

[143] *See, e.g.,* Luca Soldaini, Ai2 Dolma: 3 trillion token open corpus for language model pretraining, ALLEN AI BLOG (Aug. 18, 2023), https://allenai.org/blog/dolma-3-trillion-tokens-open-llm-corpus-9a0ff4b8da64.

[144] 2022 World Factbook Archive, Central Intelligence Agency, https://www.cia.gov/the-world-factbook/about/archives/2022/countries/world/ (last visited July 7, 2025).

[145] Zaid Zaid & Vincent Voci, *Global expansion in Generative AI: a year of growth, newcomers, and attacks*, THE CLOUDFLARE BLOG (Mar. 10, 2025), https://blog.cloudflare.com/global-expansion-in-generative-ai-a-year-of-growth-newcomers-and-attacks/.

[146] Democratization probably also requires sharing "negative knowledge," which are lessons learned from what didn't work, to guide future research and help others avoid similar pitfalls.

[147] For example, ask a college-educated layperson to read *2 OLMo 2 Furious* and explain what's going on. They almost certainly won't be able to explain even a third of the paper.

Unsurprisingly, relatively few people utilize even the most comprehensive free fine-tuning resources, such as Ai2's Tulu 3 collection, which includes training repositories, evaluation repositories, datasets, models, a demo, and an explanatory technical paper.[148]

## Elements Necessary for Democratization of GenAI:

The most helpful framework for evaluating open-source GenAI consists of the fourteen elements discussed in Rethinking Open Source Generative AI[149]:

| | **Component** | **Description** |
|---|---|---|
| 1 | Base Model Data | Are data sources for training the base model comprehensively documented and freely made available? If a distinction between base (foundation) and end (user) models is not applicable, this mirrors the end-user model data entries. |
| 2 | End User Model Data | Are data sources for training the model that the end user interacts with, including fine-tuning data, comprehensively documented and made available? |
| 3 | Base Model Weights | Are the weights of the base models made freely available? If a distinction between base (foundation) and end (user) models is not applicable, this mirrors the end-user model data entries. |
| 4 | End User Model Weights | Are the weights of the model that the end user interacts with made freely available? |
| 5 | Training Code | Is the source code of the dataset processing, model training, and tuning comprehensively made available? |
| 6 | Code Documentation | Is the source code for the data source processing, model training, and tuning comprehensively documented? |
| 7 | Hardware Architecture | Is the hardware architecture used for data source processing and model training comprehensively documented? |
| 8 | Preprint | Are archived preprints(s) available that detail all major parts of the system, including data source processing, model training, and tuning steps? |

[148] Hugging Face, *Llama-3.1-Tulu-3-8B*, https://huggingface.co/allenai/Llama-3.1-Tulu-3-8B (last visited July 7, 2025).

[149] Andreas Liesenfeld & Mark Dingemanse, *Rethinking open source generative AI: open-washing and the EU AI Act*, FACCT 1774, 1774-1787 (2024), https://dl.acm.org/doi/10.1145/3630106.3659005.

| 9 | Paper | Are peer-reviewed scientific publications available that detail all major parts of the system, including data source processing, model training, and tuning steps? |
|---|---|---|
| 10 | Model Card | Is a model card in a standardized format available that provides comprehensive insight into model architecture, training, fine-tuning, and evaluation? |
| 11 | Datasheet | Is a datasheet, as defined in "Datasheets for Datasets" (Gebru et al., 2021), available? |
| 12 | Package | Is a packaged release of the model available on a software repository (e.g., a Python Package Index, Homebrew)? |
| 13 | API and Meta Prompts | Is an API available that provides unrestricted access to the model (other than security and CDN restrictions)? If applicable, this entry also collects information on the use and availability of meta prompts. |
| 14 | Licenses | Is the project fully covered by Open Source Initiative-approved licenses, including all data sources and training pipeline code? |

However, these fourteen elements address only the traditional artifacts that enable access; they do not encompass the additional enablers required for true democratization. Even if an entity releases data, code, weights, or related artifacts (or any combination thereof), it rarely provides the necessary GPUs, engineering infrastructure, and talent to use them effectively. Without this complete package, leveraging the components above becomes prohibitively complex and costly.

Each of the following components is also essential for the effective use of the artifacts above.

| Component | Description |
|---|---|
| Computational power | Training leading models requires immense, expensive computing resources. |
| Money | Substantial monetary resources are needed not only for computational power but also to sustain ongoing development and innovation. |
| A labor force | ● Data annotation (and the data itself)<br>● RLHF (and that fine-tuning data)<br>● AI “tutors” (and their prompts)<br>● Content moderation (trust and safety) |
| Expertise to know how to use all the above[150] | Technical proficiency and domain-specific knowledge are indispensable for effectively harnessing all the above. Without such expertise, even a full suite of tools and infrastructure remains underutilized.<br>● Engineers<br>● Software developers<br>● Product developers<br>● Maintenance<br>● R&D |
| A voice in decision-making | It cannot be a democratized outcome if there is no democratic process for deciding what is created, by whom, and for what purposes. |

As David Gray Widder, a postdoctoral fellow at Cornell Tech, observed, "Even maximally open A.I. systems do not allow open access to the resources necessary to 'democratize' access to A.I., or enable full scrutiny."[151]

## Human Labor

The human workforce required to produce model outputs relies heavily on precarious labor. Even if all the data created by these workers were released alongside the models, it remains unclear how this approach aligns with the principles of democratization. Exploitative labor practices, such as paying minimal wages, providing limited job security, and exposing contractors to tedious, toxic content for extended periods, more closely resemble colonialism than democracy.[152]

Satisfying the portion of democratization that requires GenAI development to be "organized or operated so that all people involved have power, influence, etc." would require transparency throughout the entire operational process, including disclosures of how data was collected, how annotators provided feedback,

---

[150] Reuters, *OpenAI, Google, xAI Battle for Superstar AI Talent, Shelling Out Millions*, May 21, 2025, https://www.reuters.com/business/openai-google-xai-battle-superstar-ai-talent-shelling-out-millions-2025-05-21/.

[151] Kalley Huang, *What Is 'Openwashing' in AI?*, N.Y. Times (May 17, 2024), https://www.nytimes.com/2024/05/17/business/what-is-openwashing-ai.html.

[152] *See, e.g.*, Josh Dzieza, *The human cost of AI*, The Verge (June 21, 2023), https://www.theverge.com/features/23764584/ai-artificial-intelligence-data-notation-labor-scale-surge-remotasks-openai-chatbots; Billy Perrigo, *Inside Facebook's African Sweatshop*, Time (Feb. 17, 2022), https://time.com/6147458/facebook-africa-content-moderation-employee-treatment/.; Billy Perrigo, *OpenAI Used Kenyan Workers Earning Less Than $2 Per Hour to Make ChatGPT Less Toxic*, Time (Jan. 18, 2023), https://time.com/6247678/openai-chatgpt-kenya-workers/.; Karen Hao and Deepa Seetharaman, *ChatGPT, OpenAI Content Abusive, Sexually Explicit Harassment, Kenya Workers on Human Worker*s, *Wall Street Journal* (last visited July 7, 2025), https://www.wsj.com/tech/chatgpt-openai-content-abusive-sexually-explicit-harassment-kenya-workers-on-human-workers-cf191483.; Christian Shepherd & Shibani Mahtani, *The Dark Side of AI: Exploiting the Poor to Train Models,* WASH. POST (Aug. 28, 2023), https://www.washingtonpost.com/world/2023/08/28/scale-ai-remotasks-philippines-artificial-intelligence/.

where human laborers were located, and how these individuals were compensated. If the system is characterized by one-sided power dynamics, exploitation, and limited upward mobility, it is unlikely to be democratized.

## Why Not Democratize GenAI?

Several factors might discourage companies from pursuing truly democratized AI systems. The reasons vary, but the primary concern is protecting trade secrets and confidential information. Trade secrets are information that satisfies three criteria: (1) the information is not generally known (hence, secret), (2) the information provides economic value (hence why one might want to keep it secret), and (3) reasonable measures are taken to maintain secrecy.

Consider reinforcement learning from human feedback (RLHF), which is what most human laborers in the preceding section contribute to. The process is extremely resource-intensive, often requiring the contributions of thousands of people worldwide and costing many millions of dollars. Yet despite its broad applicability across various models and enduring relevance, RLHF feedback typically remains proprietary, even when companies open-source other portions of their AI systems. Most open-source GenAI entities rely on a handful of off-the-shelf datasets, further homogenizing the field.

Decision-making processes within open-source organizations offer additional insight into the limitations of democratization. Unsurprisingly, leadership and decision-making roles are overwhelmingly occupied by homogeneous groups, predominantly individuals from WEIRD (Western, Educated, Industrialized, Rich, and Democratic) countries. There is minimal direct input from broader communities impacted by these decisions.[153] Their business partners, too, tend to represent exclusive, elite circles rather than local governments or public interest organizations.[154]

Despite these criticisms, and although open-source initiatives do not increase democratization to the extent advocates claim, this limitation does not negate *all* democratizing effects or render open-source inherently problematic. However, the claim that open source will lead to an ideal world of democratized artificial intelligence simply by making data, code, weights, and evaluations available is, at best, overly optimistic.

Moreover, one might question whether complete democratization of such transformative technology is entirely desirable. As *Ars Technica* observes, "With [Google's] Veo 3's ability to generate convincing video with synchronized dialogue and sound effects, we're not witnessing the birth of media deception—we're seeing its mass democratization. What once cost millions of dollars in Hollywood special effects can now be created for pocket change."[155] Safety is discussed more thoroughly below.

---

[153] *See, e.g.,* Allen Institute for AI, *About*, allenai.org, https://allenai.org/about (last visited July 7, 2025).
[154] *See*, *e.g.*, https://allenai.org/
[155] Benj Edwards, *AI Video Just Took a Startling Leap in Realism. Are We Doomed?*, ARS TECHNICA (May 29, 2025), https://arstechnica.com/ai/2025/05/ai-video-just-took-a-startling-leap-in-realism-are-we-doomed/.

## C. Environmental Impact and Open-Source GenAI

Another assertion holds that open-source GenAI models are more environmentally friendly than their closed-source counterparts. The basic premise is that open-source models spare people the need to train their own models, thereby avoiding the environmental costs of that training. However, this perspective often overlooks the practical realities of deployment and the human behavior patterns that emerge at scale.

As with other policy topics in this section, the argument here is not that open-source is especially harmful to the environment, but that it is not especially better for the environment than closed-source models, and therefore is no more deserving of regulatory leniency than closed-source providers. The analysis that follows contends that closed-source models, due to their centralized control, optimized infrastructure, and powerful economic incentives for efficiency at scale, may have a lower overall environmental footprint than the fragmented, often suboptimal, and widely proliferated nature of open-source deployments.

A key goal of open-source initiatives is to enable and encourage others to build their own models. While releasing these artifacts openly may help others understand model development and functionality, it can also produce unintended environmental consequences. This outcome directly contradicts the public statements of many open-source entities regarding environmental responsibility and their commitments to mitigating environmental harm. As GenAI becomes easier and cheaper to develop, increasing numbers of individuals and organizations will inevitably experiment with and deploy these models. In aggregate, millions of instances running on diverse hardware, from older GPUs and consumer-grade CPUs to personal laptops, result in dramatically increased compute demands, substantially higher electricity usage, and significantly greater water consumption for cooling. While fine-tuning an open model (such as Mistral or Llama) requires fewer computational resources than full pre-training, the collective environmental impact can still be substantial. If open-source entities genuinely seek to benefit humanity through their work, it's unclear how they adequately account for this environmental impact in their cost-benefit calculations.

The scale of this environmental burden becomes apparent when examining concrete data. Currently, neither open-source nor closed-source GenAI approaches come close to offsetting their carbon emissions through efficiencies elsewhere; therefore, the discussion primarily centers on which types of models emit more carbon.[156]

In one of the few comprehensive analyses on the topic, researchers from Ai2 estimate that creating their OLMo 7B[157] model emitted seventy tons of carbon dioxide.[158] They also estimate that Meta's Llama 3.1 8B produced 420 tons of carbon emissions.[159]

---

[156] While models may improve in efficiency, *reducing* carbon emissions is not the same as achieving *zero* or *negative* carbon emissions—a critical distinction often overlooked in industry discourse.

[157] The B stands for billion. OLMo 7B has seven billion parameters, which is how we compare model sizes. So, Llama 70B is ten times bigger than OLMo 7B.

[158] Pete Walsh et al., *2 OLMo 2 Furious* (Jan. 15, 2025), https://arxiv.org/pdf/2501.00656#page=27.77

[159] *Id.*

For perspective, Ai2 and Carnegie Mellon University researchers note that OLMo 7B's energy usage would have been enough to power an average U.S. home for 13 years and 6 months.[160] Llama 3.1 8B required enough energy to power a house for 83 years.[161]

It's important to note that these are relatively small models and reflect only those released (i.e., they exclude failed runs or underperforming models that companies chose not to release). For context, Meta also offers 70B and 405B models, which required far more energy to train and emitted significantly more carbon dioxide. Currently, there is no evidence that large GenAI models are carbon-negative or will ever become so.

One of the primary weaknesses of the open-source argument is its potential for uncontrolled proliferation and redundant inference. While open-source models might reduce the need for multiple companies to redundantly train foundational models, they dramatically increase the potential for redundant inference and suboptimal deployment. A closed-source model provider typically invests enormous resources in developing highly optimized, centralized infrastructure designed for maximum efficiency per inference. This includes leveraging massive economies of scale in large data centers, which feature highly efficient power delivery, advanced cooling systems (such as liquid cooling), and specialized hardware (such as custom ASICs or the latest GPUs) that are prohibitively expensive or impractical for individual open-source users and small organizations to acquire.

Furthermore, these providers continuously fine-tune their entire inference pipeline, from network latency to model serving architecture, to minimize energy consumption per query. They utilize sophisticated load balancing and resource sharing to ensure that compute resources operate at near-peak efficiency, thereby avoiding costly idle power draw. In stark contrast, the free availability of open-source models leads to their widespread and fragmented deployment, where millions[162] of individual instances inevitably run on diverse, often inefficient hardware, usually connected to less efficient residential electricity grids. Each of these instances consumes significant idle power when not actively in use or runs on a machine not primarily dedicated to AI tasks, yet still consumes power. This pattern leads users to run models unnecessarily, experiment without clear objectives, or host redundant instances across multiple personal devices. Critically, no central entity is incentivized to optimize the collective energy consumption of these disparate open-source deployments; each user acts independently, often prioritizing ease of access or individual cost over global environmental efficiency. The aggregate effect of millions of small, inefficient deployments may ultimately exceed the environmental footprint of a limited number of large, highly optimized closed-source operations.

Additionally, open-source models inherently obscure and diffuse environmental responsibility. While the entity that trains and releases a foundational model might report its training costs, the vast majority of ongoing inference energy consumption is spread across countless users, making it largely untraceable and unreported. In practice, no mechanism exists to aggregate the energy consumption of millions of disparate open-source deployments; an individual running a model on their laptop typically does not track their specific carbon footprint from that activity. This distributed nature also makes it exceedingly difficult to regulate or implement green practices across an uncoordinated global network of individual users. Consequently, the "out of sight, out of mind" effect renders the collective environmental impact of open-source AI virtually invisible and irrelevant to any single actor, almost certainly leading to a higher overall yet unacknowledged ecological burden per user.

---

[160] Jacob Morrison et al., Holistically Evaluating the Environmental Impact of Creating Language Models (Mar. 3, 2025), https://arxiv.org/html/2503.05804v1

[161] The massive difference in energy requirements despite the small difference in model size was because Llama trained on many trillions more tokens and used an energy grid that burned more fossil fuels.

[162] As noted above, Hugging Face hosts over a million open AI models. Additionally, Meta claims its Llama models have been downloaded over a billion times. Haje Jan Kamps, *Meta says its Llama AI models have been downloaded 1.2B times*, TECHCRUNCH (Apr. 29, 2025), https://techcrunch.com/2025/04/29/meta-says-its-llama-ai-models-have-been-downloaded-1-2b-times/.

While some might argue that market forces will eventually resolve these inefficiencies, it's unclear whether this outcome is even theoretically preventable. We can only hope that the open-source entities are correct in asserting that GenAI is worth the carbon emissions, water use, electricity rate increases, environmental destruction from clear-cutting forests to build data centers, and the opportunity cost of diverting capital to such construction rather than to other pressing societal issues. However, the evidence suggests otherwise: the benefits have not offset the costs, and the trendline does not indicate they ever will.

A more responsible approach would require open-source entities to either develop a compelling justification for researching state-of-the-art general-purpose GenAI specifically, rather than focusing on other forms of AI that offer demonstrable benefits not already provided by tech giants,[163] or to switch to more focused models with clearly foreseeable benefits. GenAI not only likely commands the bulk of its developers' technical and financial resources but also likely represents the primary source of legal and reputational risks.

In summary, the theoretical energy savings from open-source models may appear viable only under highly disciplined, specialized conditions, whereas broader, unmanaged proliferation will almost certainly result in a greater net negative impact on the planet than open-source GenAI.

## D. Innovation and Open-Source GenAI

Open-source GenAI is no closer to solving meaningful problems, such as disease, poverty, or global warming, than its closed-source counterparts. Moreover, most major breakthroughs in GenAI technology development, as well as applications built on top of GenAI models developed by others, have originated in the proprietary sector rather than in open-source initiatives. This isn't to say that open-source does not lead to innovation, just that it is far from clear that it leads to *more* innovation than the proprietary sector, thereby justifying looser oversight of open-source GenAI.

Here, innovation encompasses novel developments that deliver meaningful value to society through widespread adoption and application. While the narrative of open-source GenAI as the ultimate engine of innovation and societal progress is superficially compelling, it has notable limitations and potential drawbacks that warrant serious examination. Logic suggests that closed-source models, precisely because of their centralized control, concentrated resources, and inherent accountability mechanisms, are demonstrably more effective at driving responsible, impactful innovation for broader societal benefit. The very democratization lauded in open-source discourse can inadvertently lead to the proliferation of unsafe or suboptimal applications, the diffusion of critical resources, and the fragmentation of effort, ultimately hindering rather than accelerating genuine progress.

[163] *See, e.g.,* Allen Institute for AI, EarthRanger, *available at* https://allenai.org/earthranger (last visited July 7, 2025); Allen Institute for AI, Skylight, *available at* https://allenai.org/skylight (last visited July 7, 2025); Allen Institute for AI, Wildlands, *available at* https://allenai.org/wildlands (last visited July 7, 2025); Allen Institute for AI, Climate Modeling, *available at* https://allenai.org/climate-modeling (last visited July 7, 2025).

First, true cutting-edge innovation in GenAI consistently demands immense, concentrated resources and long-term strategic vision that are almost exclusively held by large, well-funded organizations operating with a closed-source model. Ai2's CEO, Ali Farhadi, has stated his position on open-source and innovation unequivocally: "The biggest threat to AI innovation is the closed nature of the practice."[164] However, this assertion appears to be contradicted by empirical evidence. Without open source, he seems to suggest, the companies developing closed-source models that have consistently outpaced open-source ones will, for unexplained reasons, stop innovating. Yet developing foundation models requires staggering computational power, vast proprietary datasets, and access to the world's top AI talent – resources that are simply unattainable for individual developers or small open-source communities. Building on open-source models also presents challenges, including high computational costs, reliance on public datasets, limited market access, short time horizons for breakthroughs (often due to financial constraints), and the upkeep and quality control needed to maintain systems (often reliant on volunteers). These concentrated investments enable unprecedented breakthroughs, leaps in model capabilities, and the kind of foundational research that genuinely pushes the boundaries of AI. While open-source projects can fine-tune and iterate on existing models, the most significant, expensive, and often transformative leaps consistently originate in environments where resources can be aggregated and directed with a singular focus on grand challenges. While the open-source community undoubtedly benefits from advances pioneered by proprietary actors, it seldom serves as the primary source of foundational innovation.

Second, the quality, reliability, and long-term sustainability of innovation are more consistently achieved through closed-source development. Proprietary models undergo rigorous internal testing and extensive quality assurance, and are designed for robust, reliable performance in critical applications. Commercial imperatives and user expectations for stability and support drive this meticulous engineering. Open-source projects, while potentially vibrant, often face fragmentation, inconsistent maintenance, variable documentation quality, and a lack of clear support channels. For meaningful societal progress, particularly in sensitive sectors such as healthcare, infrastructure, or finance, reliability and trust are non-negotiable. Deploying a potentially unstable or poorly maintained open-source model, regardless of its purported innovation, can lead to catastrophic failures, systematically erode public trust in AI, and significantly hinder its responsible integration into society. The long-term economic incentive for closed-source developers to maintain and continually evolve their products leads to more sustained and dependable innovation cycles.

Third, while open-source GenAI theoretically promotes wide accessibility, it also introduces substantial "noise" that obscures the "signal" and can discourage foundational investment. The overwhelming volume of redundant or low-quality open-source projects makes it extremely challenging to identify genuinely transformative innovations, paradoxically slowing collective progress by overwhelming users and researchers with options that may be neither robust nor well-maintained. Furthermore, the lack of robust intellectual property protection inherent in open-source models actively discourages the colossal investments required for pioneering research and development by the very entities most capable of making profound breakthroughs. The economic logic is straightforward: why invest billions in creating a revolutionary model if competitors can immediately commoditize its core intellectual value? This dynamic leads to a focus on incremental improvements rather than bold, high-risk, high-reward foundational research that typically defines genuine societal progress. From this perspective, controlled innovation, guided by significant investment and clear accountability, will ultimately yield more stable, impactful, and trustworthy AI advancements that benefit society.

---

[164] Mark Sullivan, *Why Ai2's Ali Farhadi advocates for open-source AI models*, FAST COMPANY (Feb. 24, 2025), https://www.fastcompany.com/91283517/ai2s-ali-farhadi-advocates-for-open-source-ai-models-heres-why.

Finally, the paramount concern with open-source GenAI's unrestricted access is the heightened risk of misuse and the proliferation of potentially harmful applications, which directly undermines any claims of societal progress. While closed-source developers must grapple with the ethical deployment of their models, their centralized control enables stricter guardrails, content filtering, and robust safety protocols to prevent malicious use (e.g., creating deepfakes, generating harmful misinformation, or assisting in illicit activities). The substantial legal and reputational risks associated with large corporations deploying unsafe AI create powerful incentives to exercise greater caution and make significant investments in responsible AI development, compared with open-source projects. In contrast, open-sourcing powerful, dual-use models inevitably leads to adoption by actors with malicious intent, who systematically bypass safety measures and proliferate capabilities that sow societal discord, erode trust, or even threaten national security. This unmitigated availability means that the "innovation" occurring at the fringes is often demonstrably detrimental, requiring significant societal resources to counteract and thereby substantially hindering net progress.

## Innovation is Different from Societal Progress

We should also exercise caution when equating innovation with societal progress. These concepts are distinct. Asbestos, lead paint, slavery, CFCs, and PFAS (forever chemicals) were all innovations, yet their existence didn't mean they helped society progress or that their benefits outweighed the harms. As noted above, innovations in GenAI appear to cause far more environmental damage than they prevent, while also undermining creative workforces, consolidating market power, and facilitating the rapid spread of misinformation. Currently, there is no compelling reason to believe that the world is a better place today because of general-purpose GenAI than it was before late 2022, when ChatGPT debuted with its much-publicized technology.

## Solutionism

The response to criticisms of GenAI deployment often reveals a troubling pattern of technological solutionism.[165] Evan Selinger helpfully describes it:

Coined by the technology critic Evgeny Morozov, technological solutionism is the mistaken belief that we can make great progress in alleviating complex dilemmas, if not remedy them entirely, by reducing their core issues to simpler engineering problems. It is seductive for three reasons. First, it's psychologically reassuring. It feels good to believe that in a complicated world, tough challenges can be met easily and straightforwardly. Second, technological solutionism is financially enticing. It promises an affordable, if not cheap, silver bullet in a world with limited resources for tackling many pressing problems. Third, technological solutionism reinforces optimism about innovation—particularly the technocratic idea that engineering approaches to problem-solving are more effective than alternatives that have social and political dimensions.[166]

[165] It’s similar to arguments that MOOCs would magically erase systemic inequalities in education or predictive policing would prevent crime.

The solution to problems with GenAI is supposedly simple (and exceedingly convenient for researchers who want to open-source everything): just conduct more research. According to this view, the solution to technological problems is invariably more technology, without humanity needing to complicate things with social interactions, politics, emotions, or culture. With sufficient funding, the argument goes, we can "solve" issues of bias and inequity, just as the technology industry's light touch and tens of billions of dollars have resolved problems associated with social media, such as addiction, manipulation, decreased self-esteem, and the proliferation of surveillance capitalism.

However, this argument is overly simplistic. It overlooks the complexities of reality, treating a single approach as the best and most effective solution to problems that no breakthrough AI could ever solve. Even the most theoretically perfect GenAI, for example, would not solve housing. Housing policy must navigate NIMBY concerns, zoning requirements, manufacturing and distribution logistics, and financing, among countless other factors. Similarly, technology alone can't fix education, housing, domestic violence, food shortages, transportation, drug dependencies, healthcare, mental health, underfunded libraries, racism, or homophobia, among other pressing societal challenges.

## E. Safety and Open-Source GenAI

Open-source models pose unprecedented safety risks. While closed-source platforms like ChatGPT, Claude, and Gemini can rapidly roll out patches and filters to curb misuse, open-source safeguards can be trivially removed, leaving models highly vulnerable to abuse. Unlike proprietary systems, which can immediately cut off access to those who abuse the platforms, open-source entities lose all control over their technology as soon as any user downloads it. As a result, the potential for misuse, ranging from generating convincing misinformation to executing sophisticated cyberattacks, increases exponentially. This lack of gatekeeping exposes society to unprecedented risks when powerful tools fall into the hands of malicious actors.

A common refrain among open-source GenAI advocates is that technological progress will ultimately resolve these safety issues. Open-source developers are characteristically quick to acknowledge potential harms but ultimately disclaim all responsibility for them. Hugging Face exemplifies this pattern with regard to the environment, misinformation, and numerous other concerns[167] Ai2 employs similar rhetoric, stating that although serious outcomes may result from deploying open-source AI, these problems will somehow resolve themselves if the community pursues more open-source research. This approach resembles firefighters selling flamethrowers.

[166] Kevin Driscoll, *ChatGPT Is Just Another Example of Dangerous AI Hype*, SLATE (Mar. 29, 2023), https://slate.com/technology/2023/03/chatgpt-artificial-intelligence-solutionism-hype.html.
[167] Margaret Mitchell, *A Weaponized AI Chatbot Is Flooding City Activity*, LinkedIn (Aug. 24, 2020), https://www.linkedin.com/posts/margaret-mitchell-9b13429_a-weaponized-ai-chatbot-is-flooding-city-activity-7333917467246792705-

A quote from TechCrunch illustrates this perspective:

> There has been some debate recently over the safety of open models, with Llama models reportedly being used by Chinese researchers to develop defense tools. When I asked Ai2 engineer Dirk Groeneveld in February whether he was concerned about OLMo being abused, he said that he believes the benefits ultimately outweigh the harms.
>
> "Yes, it's possible open models may be used inappropriately or for unintended purposes," he said. "[However, this] approach also promotes technical advancements that lead to more ethical models; is a prerequisite for verification and reproducibility, as these can only be achieved with access to the full stack; and reduces a growing concentration of power, creating more equitable access."[168]

Among certain open-source advocates, the optimistic view holds that technological progress will continually yield positive outcomes, enabling researchers to continually refine model safety. However, this optimism applies only to the initially released models. Once a model is in the public domain, anyone can modify it, including in ways that systematically undermine its ethical safeguards. The insights gained from open-source work can be harnessed just as readily to create harmful GenAI. This reality is not a one-way ratchet; it is far from certain that the benefits will consistently outweigh the potential harms.

The most significant safety concern with open-source GenAI is the unmitigated risk of malicious use and the potential for widespread proliferation of dangerous capabilities. By indiscriminately distributing sophisticated AI models, open-source initiatives inadvertently empower both legitimate developers and those with harmful intentions. While legitimate researchers and developers may benefit, malicious actors can simultaneously generate highly convincing misinformation, execute sophisticated cyberattacks, create hyper-realistic deepfakes for fraud or harassment, or even aid in developing harmful biological or chemical agents. Unlike closed-source providers, who can implement strict access controls, user vetting, and robust content moderation to prevent such abuses, open-source models inherently lack these gatekeeping

---

qutI?utm_source=share&utm_medium=member_desktop&rcm=ACoAABW04k4BX6b55vEcea6IlS9dOJXgjjeUCfQ. 2025, ("Things we foresaw and warned about #43920: Weaponized AI chatbot for en masse misinformation, aimed to influence policy in a way that hurts people. Let's nip this in the bud before it has more serious lasting effects!"); Sasha Luccioni, *How Much Energy Does a State-of-the-Art Video . . .*, LinkedIn (Oct. 18, 2023), https://www.linkedin.com/posts/sashaluccioniphd_how-much-energy-does-a-state-of-the-art-video-activity-7346196411362779136-Y19G?utm_source=share&utm_medium=member_desktop&rcm=ACoAABW04k4BX6b55vEcea6IlS9dOJXgjjeUCfQ. ("How much energy does a state-of-the-art video generation AI model use? 📺
A 14B parameter model like WAN2.1 uses ~109 Wh, or the equivalent of 7 full smartphone charges, for a single video! 😳")

[168] Kyle Wiggers, *AI2 Releases New Language Models Competitive with Meta's LLaMA*, TechCrunch (Nov. 26, 2024), https://techcrunch.com/2024/11/26/ai2-releases-new-language-models-competitive-with-metas-llama/.

mechanisms, making it orders of magnitude easier to develop and deploy dangerous applications with minimal oversight or consequence. From this perspective, radical accessibility constitutes a profound safety liability for society.

Furthermore, closed-source models consistently benefit from centralized guardrails, rigorous pre-release safety testing, and a chain of accountability largely absent in the open-source ecosystem. Major closed-source developers invest enormous resources in red-teaming and continuous monitoring, driven by substantial legal and reputational liability. They have a strong economic interest in ensuring their models are safe and reliable before and after deployment. They often implement proprietary safety layers and adversarial training techniques that are deliberately kept secret from potential attackers, making them significantly harder to circumvent. In contrast, while open-source projects can identify certain bugs, no mechanism exists to ensure that fixes are universally or consistently applied across all users' deployments. Users routinely run outdated, unpatched, or poorly secured versions, and the ecosystem's fragmentation makes it nearly impossible to enforce safety standards or quickly mandate widespread updates when new vulnerabilities are discovered. The responsibility for safe deployment falls entirely on individual users, who frequently lack the expertise, resources, or even the inclination to prioritize safety, resulting in a highly fragmented and inconsistent safety posture across the broader open-source landscape.

Some proponents, including Ai2's CEO, argue that open-source initiatives can help remedy the harms that GenAI can cause. "People will do bad things with this technology," Mr. Farhadi said, "as they have with all powerful technologies." The task for society, he added, is to better understand and manage the risks. Openness, he contends, is the best bet for finding safety and sharing economic opportunity. 'Regulation won't solve this by itself,' Mr. Farhadi said."[169] Yet it remains entirely unclear how increased awareness alone can prevent misuse, since the very same knowledge will inevitably be leveraged to amplify harmful applications.

While transparency is often cited as a safety benefit, it can also be a double-edged sword, particularly for security. Openly revealing a model's architecture, weights, and significant portions of its training methodology can inadvertently provide detailed blueprints that attackers can use to identify and exploit vulnerabilities or bypass safety features more effectively than if those internal workings remained proprietary. While closed-source models are often criticized as "black boxes," their safety mechanisms can be "black boxes" as well, benefiting from security through obscurity. Moreover, the massive resources dedicated by leading closed-source labs to cutting-edge AI safety research, alignment techniques, and robust system hardening often far surpass what fragmented open-source efforts can achieve. This concentrated expertise and financial backing enable the incorporation of more advanced, preventative safety measures into the architecture of the models, arguably leading to more secure and responsibly managed AI systems that substantially minimize broad societal risks stemming from the proliferation of powerful yet unconstrained technology.

When confronted with these safety concerns, open-source advocates often resort to whataboutism, deflecting criticism by pointing to other technologies and saying, "Well, what about [Google search]?" or by citing alternatives, thereby claiming that releasing these models does not add any new risk. But this rhetorical strategy is either misguided or disingenuous. It is akin to justifying the release of a GenAI that can quickly produce hyper-realistic video and photographic content with minimal technical skill by arguing that harmful visual content has always been accessible. Such justification amounts to willful ignorance, much like a chemical company that excuses its environmental damage by claiming that other companies have already polluted our waters. Bad actors have an asymmetrical advantage when it comes to GenAI harm: they need to succeed only rarely to be successful. A steady but occasional triumph, intended to undermine trust, destroy self-esteem, spur violence, and perpetrate fraud, is sufficient.

---

[169] Cade Metz, *An Industry Insider Drives an Open Alternative to Big Tech's A.I.*, N.Y. TIMES (Oct. 19, 2023), https://www.nytimes.com/2023/10/19/technology/allen-institute-open-source-ai.html.

What makes GenAI unique is its ability to generate novel, useful outputs efficiently, at scale, and at remarkably low cost. This includes outputs that can be used for malicious purposes. Without these capabilities, GenAI would not be the transformative technology worth spending hundreds of millions of dollars (for open-source entities) or hundreds of billions of dollars (for proprietary systems) to develop.

As David Widder and Dawn Nafus observed in their paper on accountability in the AI supply chain:
"We were struck by the deeply dislocated sense of accountability, where acknowledgement of harms was consistent, but nevertheless another person's job to address, always elsewhere."[170]

This observation highlights an issue in the open-source AI community: a systematic diffusion of responsibility that enables the continued deployment of harmful models while evading accountability.

A final point is that many open-source entities lack the robust security systems of their better-funded proprietary rivals. When deploying models, the absence of stringent security protocols can allow malicious actors to access sensitive user data or compromise the GenAI system itself. Platforms such as ChatGPT and Anthropic's services follow established cybersecurity frameworks, often attain SOC 2 Type 2 certification, and invest dedicated budgets and personnel in security. In contrast, open-source projects often operate without access to a dedicated information security team or a comprehensive security budget, despite their risks being essentially identical to those of companies like OpenAI. Crucially, information privacy cannot exist without robust information security.

## F. Public Policy Conclusion

The review of public policy arguments has highlighted both the strengths and limitations of open-source GenAI, showing that its purported benefits are often overstated or unproven. Even when policy goals are well-intentioned, open-source initiatives face unique complications that threaten legal compliance and the integrity of the movement.

[170] David Gray Widder & Dawn Nafus, *Dislocated accountabilities in the "AI supply chain": Modularity and developers' notions of responsibility*, Big Data & Society, 1–12 (June 15, 2023), https://doi.org/10.1177/20539517231177620.

# VII. Complications from Open-Source GenAI

Beyond traditional legal and policy criticisms of special exemptions for open-source GenAI projects, these initiatives face distinctive challenges that threaten both legal compliance and the integrity of the open-source movement. Some entities falsely claim open-source status, apply licenses to models incompatible with AI architectures, exploit open-source data for commercial gain, or fail to honor opt-out requests regarding the use of copyrighted or personal information. These practices not only undermine legitimate open-source efforts but also create a legal and ethical minefield.

## A. Openwashing and the Erosion of Open-Source Integrity

Openwashing—a practice whereby entities claim open-source leadership without fully embracing the principles of openness—poses a significant challenge in the GenAI landscape. Media amplification of such claims can mislead stakeholders and erode the integrity of the open-source movement. This trend is evident not only in general outlets such as the Wall Street Journal, Axios, Fortune, Forbes, and CNBC, but also in tech-focused publications such as VentureBeat and Wired.[171]

The paradigmatic example of this phenomenon is Meta, which claims to follow open-source principles by releasing its model weights but withholds elements such as training data and alignment methodologies (e.g., RLHF). This selective sharing has drawn accusations of "openwashing," a term that criticizes superficial claims of transparency without corresponding action.[172] Much like the greenwashing companies that espouse pro-environmental values while acting otherwise, organizations like Meta leverage the "open-source" label to boost their public image and dodge scientific, legal, and regulatory scrutiny, all without committing to genuine openness. Elon Musk has made a similar openwashing claim about OpenAI, arguing that OpenAI no longer adheres to its original mission of making its work open, and therefore, Musk was bamboozled by giving OpenAI tens of millions of dollars for that purpose.[173] Despite its name, OpenAI is far less open than Meta.

---

[171] *Mistral AI Drops New Open Source Model That Outperforms GPT-4o Mini with Fraction of Parameters*, VentureBeat (June 28, 2024), https://venturebeat.com/ai/mistral-ai-drops-new-open-source-model-that-outperforms-gpt-4o-mini-with-fraction-of-parameters/.;
Mauro Orru, *Mistral AI Bets on Open-Source Development to Overtake DeepSeek, CEO Says*, Wall St. J. (Mar. 7, 2025), https://www.wsj.com/tech/ai/mistral-ai-bets-on-open-source-development-to-overtake-deepseek-ceo-says-de031411?gaa_at=eafs&gaa_n=ASWzDAipDxzs2P-rOoS5tcRS4V8r2Robcv-0iic7OoNOhMYGLFrcVZBRmPGOjf_5xo%3D&gaa_ts=68530424&gaa_sig=x9p20CdbhXVe06Hkrx9E5rNNfsBJywkY-5D6iR58-gUVQ2qjx5pHTyt-fxtk1uIkdqtMRDG4DjWYL7nvVGO8bQ%3D%3D.;Axios, *How Open Source AI Is Leveling the Playing Field*, https://www.axios.com/sponsored/how-open-source-ai-is-leveling-the-playing-field; Jeremy Kahn, *Deepseek Just Flipped the AI Script in Favor of Open Source—And the Irony for OpenAI and Anthropic Is Brutal*, Fortune (Jan. 27, 2025), https://fortune.com/2025/01/27/deepseek-just-flipped-the-ai-script-in-favor-of-open-source-and-the-irony-for-openai-and-anthropic-is-brutal/; Will Knight, *Meta's Open Source Llama 3 Is Nipping at OpenAI's Heels*, Wired (Apr. 18, 2025), https://www.wired.com/story/metas-open-source-llama-3-nipping-at-openais-heels/; Ryan Browne, *Deepseek Breakthrough Emboldens Open-Source AI Models Like Meta's Llama*, CNBC (Feb. 4, 2025), https://www.cnbc.com/2025/02/04/deepseek-breakthrough-emboldens-open-source-ai-models-like-meta-llama.html; Luis Romero, *ChatGPT, Deepseek, or Llama? Meta's LeCun Says Open Source Is the Key*, Forbes (Jan. 27, 2025), https://www.forbes.com/sites/luisromero/2025/01/27/chatgpt-deepseek-or-llama-metas-lecun-says-open-source-is-the-key/.

[172] The origin of the term openwashing appears to be from Michelle Thornet in 2009: Michelle Thorne, *Openwashing*, michellethorne.cc (Mar. 9, 2009), https://michellethorne.cc/2009/03/openwashing/.

Andreas Liesenfeld and Mark Dingemanse's empirical analysis reveals the depth of this transparency deficit. A comprehensive evaluation of 14 criteria (each classified as fully open, partially open, or closed) across more than 40 large language models and 10 text-to-image models shows that Meta's Llama 4 meets only three criteria fully, is partially open on four, and remains closed on seven. These criteria include access to base model data, model weights, training code, documentation, hardware details, architectural information, preprints, model cards, datasheets, packaging, API and meta prompts, and licensing, as more fully described in the Democratization section of this paper. For comparison, OpenAI's ChatGPT, which never claimed to be open source, is partially open only in terms of its preprint and closed on thirteen other measures, while the OLMo model by Ai2 is fully open on thirteen criteria and partially open on one.

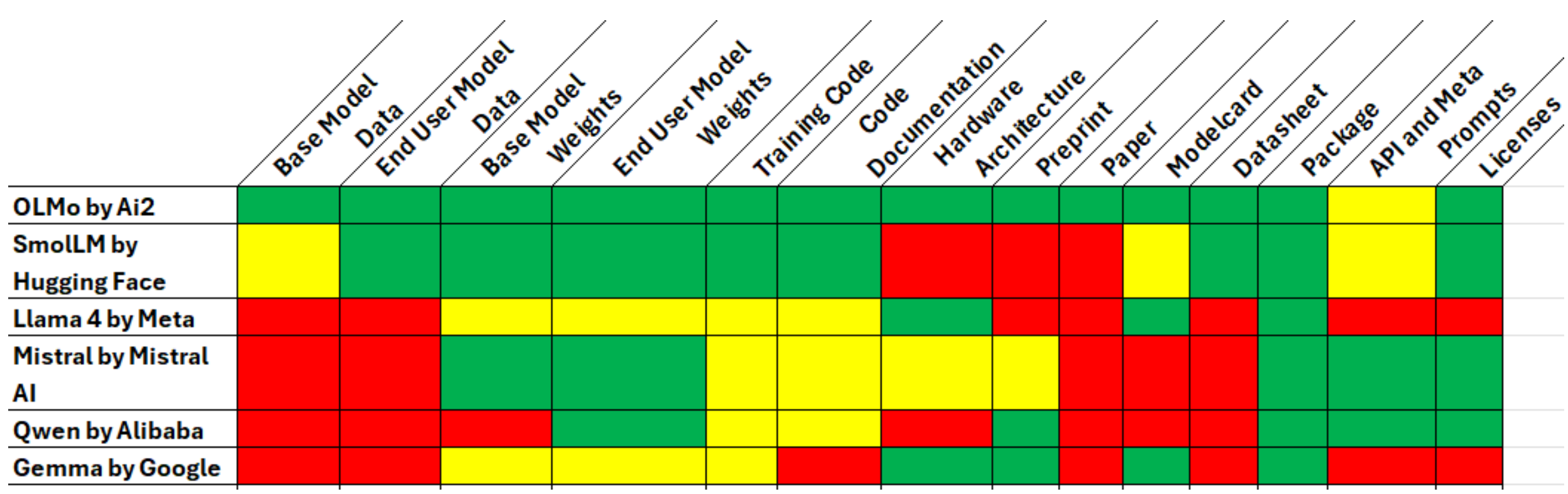

[174]

A key concern, as articulated by the Linux Foundation, is that "this 'openwashing' trend threatens to undermine the very premise of openness — the free sharing of knowledge to enable inspection, replication and collective advancement."[175] Put simply, if we keep lowering the bar for what qualifies as open source, the term will become meaningless.

Openwashing also undermines the credibility of model evaluation. When companies withhold essential components needed for replication and bypass peer review, they can selectively report benchmark results without enabling independent verification. In reality, these metrics can be artificially inflated through tailored fine-tuning, where improved scores may reflect training for a specific task rather than genuine, generalizable intelligence, and through training on test data and cherry-picking optimal results from methodologies such as zero-shot, few-shot, 25-shot, or chain-of-thought evaluations. In essence, the benchmark numbers shared in non-peer-reviewed documents by companies that do not reveal the training or fine-tuning data are hardly more informative than pure speculation. Despite this lack of scientific rigor, they author these documents in a manner that appears to convey serious scientific thought, formatting them in styles common in academia and employing academic jargon to resemble robust scientific findings.

[173] [CITE LAWSUIT]; From X: "OpenAI was created as an open source (which is why I named it "Open" AI), a non-profit company to serve as a counterweight to Google, but now it has become a closed source, maximum-profit company effectively controlled by Microsoft. Not what I intended at all."

[174] The Open Source AI Index, https://open-sourceai-index.eu/the-index?type=text&view=grid https://osai-index.eu/the-index?type=text&view=grid.

[175] Vincent Caldeira, Vincent Ng, Michael W. Stewart & Dr. Ibrahim Haddad, Introducing the Model Openness Framework: Promoting Completeness and Openness for Reproducibility, Transparency, and Usability in AI, LF AI & DATA Blog (Apr. 17, 2024), https://lfaidata.foundation/blog/2024/04/17/introducing-the-model-openness-framework-promoting-completeness-and-openness-for-reproducibility-transparency-and-usability-in-ai/.

Perhaps the most pernicious effect of openwashing is that the appropriation of the term "open source" allows companies to reap the reputational benefits of the open-source label without contributing to the community in ways that sustain and advance the open-source ethos. This practice creates a parasitic relationship: it dilutes the value of "open source," reducing it to an extractive, hollow shell of its original promise.

## B. GenAI Open-Source Licensing Challenges

Open source is not only about access to the code, data, weights, and so forth; it is equally about the terms under which these resources are shared. This further complicates matters for Meta because, although it claims to be open-source, it does not release its models under an OSI-approved license. Instead, it uses a proprietary one developed by Meta's lawyers. While the license is not restrictive for most uses, it nonetheless imposes constraints that many in the open-source community chafe at. For example, the Llama license explicitly says that "You agree you will not use, or allow others to use, Llama 4 to: …Engage in any action, or facilitate any action, to intentionally circumvent or remove usage restrictions or other safety measures, or to enable functionality disabled by Meta."[176] In other words, users who download the model are contractually prohibited from jailbreaking it for any reason, including scientific research.

The licensing landscape becomes even more complex when we consider that there is currently no widely accepted open-source license tailored specifically for GenAI. Instead, most open-source initiatives release models under conventional software licenses, such as the Apache 2.0 license. However, licenses like Apache 2.0 and MIT are designed for source code, while AI models are often distributed as binary files and training weights (essentially configuration files). The focus on source code in these licenses does not directly address the sharing and modification of AI models.

Traditional software licenses pose several challenges when applied to AI models. Apache licenses often require attribution and disclosure of changes; however, applying these requirements to AI models can be difficult. AI model modifications often involve fine-tuning the model's parameters, which may not be as straightforward as "modified code" in traditional software. Moreover, AI models rely heavily on training data. These licenses may fall short in addressing the complex issues surrounding the sharing and licensing of this data, particularly privacy and copyright considerations.

In addition, existing open-source licenses may not fully capture the specific nuances of AI models, including their potential for harm or misuse. Most traditional software serves a single purpose for which it was specifically designed. This is a far cry from GenAI, where the key differentiator is that GenAI models are general-purpose. For instance, text-based models can write thank-you notes just as easily as they can compose poetry. Traditional software could write one or the other, but not both, and definitely not also explain the intricacies and plot holes of something like the show *Game of Thrones*. This difference in capability and scope demands a corresponding evolution in licensing frameworks.

[176] Llama, Llama 4 Use Policy, https://www.llama.com/llama4/use-policy/ (last visited July 7, 2025).

There is perhaps an even bigger issue with licensing models: it is unclear whether the models are copyrightable. If they are not copyrightable, licenses do not apply, because licenses are a means of specifying what can and cannot be done with copyrighted material. Instead, the best any entity can do is control what people can do at the time of download. The licenses cannot bind downstream uses. While licenses have always been a relatively weak form of enforcement for open-source technology, enforcing them for open-source GenAI models may be virtually impossible.

## C. Data Laundering

Data laundering, also called data washing, mirrors the criminal process of money laundering. According to the U.S. Treasury's Financial Crimes Enforcement Network (FinCen), "Money laundering involves disguising financial assets so they can be used without detection of the illegal activity that produced them. Through money laundering, the criminal transforms the monetary proceeds derived from criminal activity into funds with an apparently legal source."[177] Data laundering operates on the same principles: taking data from one source and repackaging it to make it appear perfectly legal for unrestricted use.[178]

Data repackaging raises several legal issues, including potential breach of contract and copyright infringement. GenAI entities assume these legal risks for various reasons, often arguing that (i) complying with the law is too cumbersome, (ii) complying with the law is too inconvenient, and (iii) the entity's use of the data outweighs the rights of the data's owner. These flimsy justifications lead to several prominent and legally problematic outcomes.

First, there is the release of datasets under the ODC-BY license, which proponents claim makes the data freely available. In reality, ODC-BY merely aggregates various licenses under a single label, effectively issuing an implied "caveat emptor" warning.[179] Users must still comply with the specific license terms

[177] FinCEN, *What is Money Laundering?*, https://www.fincen.gov/what-money-laundering (last visited July 7, 2025).

[178] Of course, the primary intent of money laundering is to conceal illicit activities, which is a crime. In contrast, data laundering may only obfuscate the data source incidentally. Still, the underlying concept is similar from a civil standpoint.

[179] The Preamble states that "Databases can contain a wide variety of types of content (images, audiovisual material, and sounds all in the same database, for example), and so this license only governs the rights over the Database, and not the contents of the Database individually. Licensors may therefore wish to use this license together with another license for the contents. Sometimes the contents of a database, or the database itself, can be covered by other rights not addressed here (such as private contracts, trademark over the name, or privacy rights / data protection rights over information in the contents), and so you are advised that you may have to consult other documents or clear other rights before doing activities not covered by this License."

Section 2.4 of the ODC-BY license further states that "The individual items of the Contents contained in this Database may be covered by other rights, including copyright, patent, data protection, privacy, or attached to each piece of content, meaning that despite its marketing, ODC-BY does not confer the unrestricted freedoms of true open-source data. Typically, the content within ODC-BY datasets remains copyrighted, and the original owners have not consented to allow their content to be reproduced, distributed, or modified into derivative works, which are three of the exclusive rights granted to copyright owners by the Copyright Act.[180] Nevertheless, companies that describe themselves as "truly open" or in similar terms often apply the ODC-BY license to their datasets, creating a misleading impression that the data can be freely used without restrictions. The marketing language obscures the fact that ODC-BY does *not* make the data openly available.

Second, some entities unilaterally change the licensing of the underlying content rather than leaving it unchanged and adding an ODC-BY layer on top. For instance, a group of researchers well-versed in software licensing took content scraped from the web and declared it open source by relicensing it under CC-BY-4.0, disregarding any existing contractual or copyright restrictions.[181] The dataset is part of DataComp-LM and consists of 240 *trillion* tokens.[182] The researchers extracted content from datasets created by Common Crawl, a nonprofit that makes "wholesale extraction, transformation and analysis of open web data accessible to researchers [and everyone else, for free and for commercial use]"[183] by scraping the internet and compiling the datasets. Rather than use ODC-BY, the researchers chose to claim that all of the content is now, through their unilateral declaration, transformed into CC-BY-4.0, which is an *actual* open-source license. In effect, they asserted that the original content creators had forfeited their exclusive copyrights. Unsurprisingly, many, if not all, of the institutions with which these researchers are affiliated have subsequently used this dataset to train and release models.

Unfortunately, these occurrences are not outliers. The Data Provenance Initiative, which is a cohort of researchers from schools like MIT and Harvard, notes that "We [] observe frequent miscategorization of licenses on widely used dataset hosting sites, with license omission of 70%+ and error rates of 50%+. This

---

personality rights, and this License does not cover any rights (other than Database Rights or in contract) in individual Contents contained in the Database. For example, if used on a Database of images (the Contents), this License would not apply to copyright over individual images, which could have their own separate licenses, or one single license covering all of the rights over the images." Open Data Commons Attribution License, v. 1.0, opendatacommons.org/licenses/by/1-0/.

[180] 17 U.S.C. § 106.

[181] The group included computer scientists from the University of Washington, Apple, Toyota Research Institute, UT-Austin, Tel Aviv University, Columbia University, Stanford, UC-Los Angeles, JSC, LAION, Ai2, Technical University of Munich, Carnegie Mellon University, Hebrew University, SambaNova, Cornell, University of Southern California, Harvard, UC-Santa Barabara, SynthLabs, Bespokelabs.AI, Contextual AI, and DatologyAI.

[182] Jeffrey Li et al., *DataComp-LM: In search of the next generation of training sets for language models*, https://arxiv.org/html/2406.11794v4

[183] CommonCrawl, https://commoncrawl.org/. Notably, Common Crawl has a proprietary license that echo pen-source ODC-BY, stating that "BY USING THE CRAWLED CONTENT, YOU AGREE TO RESPECT THE COPYRIGHTS AND OTHER APPLICABLE RIGHTS OF THIRD PARTIES IN AND TO THE MATERIAL CONTAINED THEREIN." https://commoncrawl.org/terms-of-use

points to a crisis in misattribution and informed use of the most popular datasets driving many recent breakthroughs."[184]

Third, some entities simply ignore the underlying license terms entirely. This is allegedly the case in *Doe 1 v. GitHub*, where programmers who shared code under open-source licenses such as Apache 2.0 and MIT found that their code was scraped and used to train GenAI models without regard for the specified license conditions.[185] GenAI entities, including open-source entities, typically rely on datasets such as The Stack or StarCoder, which are comprised of code files with the same types of licenses at issue in the GitHub litigation, to train models for code generation. Put simply, without these datasets and without scraping GitHub, the models would not be proficient at coding.

Importantly, some business models rely heavily on this data washing. A prime example is Hugging Face, a well-funded for-profit company that hosts a vast array of open-source artifacts (including datasets) and also creates its own. As a sophisticated player in the open-source ecosystem, Hugging Face is almost certainly aware that artifacts on its platform can facilitate the laundering of copyrighted and contract-protected works, particularly in datasets that receive hundreds of thousands or even millions of downloads. Yet the company makes no effort to require users to relicense or remove misclassified works (as seen with DCLM and its CC-BY license). This willful blindness likely stems from economic incentives: Hugging Face requires users to visit its site to download those artifacts, creating an opportunity to upsell products and services.

Another prime example is Common Crawl. Though the website claims that "We make wholesale extraction, transformation, and analysis of open web data accessible to researchers,"[186] it does not limit access to researchers. Indeed, anyone can use the data, and many people do. Common Crawl is a common source of training data for large language models, including those developed by for-profit companies. Yet websites that block OpenAI's bots and those of other prominent model developers from scraping their pages often do not block Common Crawl. This is likely because Common Crawl is not well known outside a relatively small group of researchers and developers, but it might also be because Common Crawl is a 501(c)(3) nonprofit that emphasizes its alleged research focus. As a result, companies may claim they technically follow robots.txt or other requests not to scrape website data, yet they can easily acquire that data by downloading the billions of webpages Common Crawl scrapes each month.

A further paradox emerges among staunch open-source advocates. While they may be quick to call out entities like Meta for openwashing, they are willing to ignore data washing. This selective criticism raises questions about intellectual consistency and suggests that the behavior may be hypocritical or at least questionable. For example, Ai2's CEO has remarked on Llama models, "'We love them, we celebrate them, we cherish them. They are taking the right steps. But they are just not open source.'"[187] Meanwhile, Ai2 has trained its most prized models on datasets licensed under CC-BY despite knowing that the content is not actually CC-BY, and Ai2 releases its datasets under ODC-BY, fully aware that the owners of the underlying copyrighted works have not consented to reproduction, distribution, or the creation of derivative works, and knowing that when others download ODC-BY datasets, they are almost certainly not checking the copyright status of all underlying content.

---

[184] Shayne Longpre et al., *The Data Provenance Initiative: A Large Scale Audit of Dataset Licensing & Attribution in AI* (Nov. 4, 2023) https://arxiv.org/pdf/2310.16787

[185] DOE 1 v. GitHub, Inc., No. 4:22-cv-06823 (N.D. Cal. Feb. 11, 2025) https://www.courtlistener.com/docket/65669506/doe-1-v-github-inc/

[186] https://commoncrawl.org/

[187] *Meta is accused of bullying the open-source community*, THE ECONOMIST (Aug. 28, 2024), https://www.economist.com/business/2024/08/28/meta-is-accused-of-bullying-the-open-source-community.

## D. Open Source and Opt-Out Mechanisms

Some people object to GenAI models being trained on their content, often signaling their preference through website terms of use or similar mechanisms. They rightly believe they should not have to complete a form or jump through bureaucratic hoops to protect their privacy or copyright. Nevertheless, GenAI companies use opt-out as a shield, implying that if a user fails to submit the proper form, they have effectively waived their contractual, copyright, and privacy rights.

The practical reality, however, often falls short of these promises. Even when opt-out options are announced, companies often do not follow through. For instance, in May 2024, OpenAI promised that users could specify how, or even whether, their work would be used for training. But by January 2025, that feature had yet to materialize.[188] The delay is not due to the impossibility of excluding data from datasets, but rather to the lack of incentive to do so until a court weighs in on the matter.

Even when a person opts out (notably, this is not an action required by law), that choice would, at most, apply to future model training by that single company. Such a choice would not extend to other GenAI companies, nor would it retroactively cleanse existing or previously trained models. Deleting data from datasets is only truly effective if it happens before any model training begins. Once training has commenced, the model developers have already benefited from that data, and its removal afterward won't negate that advantage. Furthermore, there's a lack of clarity about exactly what gets deleted. The original PDFs or JPEGs? The plain text extracted from those files? Or does deletion extend to the tokens generated from that text, or even the token IDs assigned to each token? Finally, the responsibility for identifying data to be deleted typically falls on the content creators, and only exact matches are likely to be removed. There's also usually no confirmation of what, if anything, was deleted, making verification impossible.

The open-source nature of the artifacts further complicates the problem. While OpenAI, Google, Anthropic, and other closed-model providers can control and modify their datasets, open-source providers cannot. Once an entity like Ai2 releases a dataset and others download it, the entity cannot claw it back. It's forever out in the open. Thus, a person may ask Ai2 to delete their information from Ai2's dataset[189], but that does not mean the dataset with the information helpfully collected and cleaned by Ai2 won't be used by other GenAI developers. In this regard, open source operates as a one-way ratchet: it can only add to others' training resources; it can't reduce them.

---

[188] Kyle Wiggers, *OpenAI failed to deliver the opt-out tool it promised by 2025*, TECHCRUNCH (Jan. 1, 2025), https://techcrunch.com/2025/01/01/openai-failed-to-deliver-the-opt-out-tool-it-promised-by-2025/.

[189] It's not clear how they'd do so. There does not appear to be an opt-out form on their dataset webpage, and searching for it on Google does not yield any results.

Moreover, open-source providers cannot control what their models do once they are released. Meta, Hugging Face, and Ai2, for example, cannot retroactively impose new filters on models that users have already downloaded. Even if they add a safeguard to the model before releasing it, these safeguards are notoriously easy to remove.[190]

These technical limitations will likely become legal shields as open-source providers argue that it's too complicated or even impossible to comply with any regulations that allow for opt-out or address violations of copyright, contract, or privacy. However, difficulty is not an affirmative defense for breaking the law. If Meta releases a model that contains a memorized copy of *Harry Potter* (which it did[191]), every single download of that model was likely a distribution of *Harry Potter* by Meta. Meta cannot now argue that it may have been a whoopsie, but because it's unable to correct it, the problem should be ignored. The same principle applies to all other open-source releases of datasets or models that contain information gathered unlawfully.

## E. Complications Conclusion

The preceding analysis has exposed a range of complications that undermine the credibility and effectiveness of open-source GenAI, from misrepresentations of openness to problematic licensing and data practices. These issues become even more pronounced during model deployment and release, when ethical and practical dilemmas arise. The following section examines these deployment challenges, focusing on the risks of mass adoption and the tension between openness and responsibility.

# VIII. Problems with Deployment and Release

There is some mission creep in open-source GenAI. Arguments that may have passed the giggle test for a purely open-source approach to conventional software do not hold when GenAI entities insist that releasing open-source models is not enough – they must also deploy them. Releasing means the files are hosted somewhere (e.g., Hugging Face), and anyone can download them, whereas deployment means the entity maintains control over the service (e.g., ChatGPT).

## A. Deployment

Open-source entities that deploy models may justify doing so by arguing that mass adoption (i.e., use) is necessary to collect and analyze enough user data to improve their models. For advocates of "truly open AI," such as Ai2, this also means making their user data public, because otherwise they'd fail to be truly open; they'd be no different from OpenAI, which collects and analyzes user data to improve its models. Moreover, sharing the data is said to underpin scientific reproducibility; however, this raises serious privacy concerns, as few users may fully appreciate that their interactions could eventually be publicly disclosed, revealing sensitive or embarrassing information.

---

[190] Pedro Cisneros-Velarde, *Bypassing Safety Guardrails in LLMs Using Humor* (Apr. 9, 2025), https://arxiv.org/html/2504.06577v1; Dmitrii Volkov, *Badllama: removing safety finetuning from Llama 3 in minutes* (July 1, 2024) (https://arxiv.org/html/2407.01376v1

[191] Feder Cooper et al., *Extracting memorized pieces of (copyrighted) books from open-weight language model*s (May 18, 2025)https://arxiv.org/abs/2505.12546

Beyond the need for data, the argument extends to commercial use. Open-source advocates argue that allowing commercial applications to use their models via APIs is the only viable path to mass adoption. This strategy not only broadens the user base but also purportedly provides developers with insights into diverse interactions, including information that can be used to create more sophisticated products for personal and consumer markets. However, critics might reasonably question whether leveraging openness for commercial viability aligns with the core principles of scientific research and public benefit.

The underlying logic holds that building merely capable models for scientific research is insufficient. Instead, it calls for developers to push toward state-of-the-art systems, meaning models that are not only highly capable but also continually evolve in response to real-world use. This shift marks a change: the technology is no longer an end in itself for basic research but rather a tool to unlock better products and services. When this rationale for massive data collection is extended to bolster AI safety, it paves the way for a surveillance model built on endless data harvesting. In this scenario, each new use case or challenge becomes a pretext for additional data collection, potentially leading to perpetual mission creep.

Eventually, this line of reasoning begins to mirror the strategies employed by tech giants such as Google and OpenAI, departing sharply from the research-first ethos traditionally cherished by universities and independent research institutes. It is difficult to envision how the reasoning would reach a point where the open-source entity would say, "Okay, we've collected enough data; now we can shift to studying it and using it for research, and we don't need to try to collect more."

The asymmetric nature of AI risks further undermines any justification based on limited data collection. Malicious actors need only a tiny fraction of the nearly limitless, low-cost outputs from large language models, such as phishing emails, misinformation, deepfakes, voice clones, and malware, to cause significant harm. In contrast, responsible developers must always prevent potential misuse. Given that it's impossible to prevent all harms from LLMs, suggesting that an open-source entity must collect ever more data to avoid harm seems not only misleading but also potentially counterproductive. In the absence of regulatory intervention, these models may already be as effective as they will ever be, as measured by the ratio of harmful outputs to overall uses.

## B. Release

The push for mass adoption also leads open-source entities to insist that they can release their models only under permissive licenses, such as the Apache 2.0 and MIT licenses. However, if the focus of open-source is truly centered on innovation and research, it's not clear why an open-source license is necessarily better than a RAIL or ImpACT license, or other licenses that impose only minimal restrictions on AI use, rather than the broad freedoms that traditional open-source licenses offer. How many researchers will be deterred from using a high-quality open-source GenAI system because they are disallowed from using it for military or commercial purposes? Releasing it under a purely permissive open-source license seems to indicate a complete indifference to how the technology is used or by whom. The ImpACT license, created by Ai2, was designed to address this shortcoming. One element required users to submit risk reports when using the license, so Ai2 would know who was using it and for what. It also required a rough estimate of environmental impact and allowed Ai2 to name and shame users who misused the AI artifacts by posting their names on a webpage and encouraging community policing.

Prioritizing rapid deployment and minimal restrictions over ethical deliberation results in what may be termed “ethical debt.” This concept captures the long-term societal costs incurred when developers neglect the ethical implications and potential harms of GenAI systems. It is a "debt" incurred by failing to address ethical issues proactively, leading to negative consequences that often affect individuals or groups with no direct connection to the product or service. Ethical debt is like technical debt in that it represents the long-term cost of short-term gains. While technical debt involves shortcuts in code that may lead to future technical difficulties, ethical debt arises from neglecting ethical considerations during the development, deployment, or use of a system. Unlike technical debt, which can sometimes be addressed by fixing the code, the consequences of ethical debt are often far more challenging to repair, especially when harm has already been done. In many cases, it's a debt that cannot be repaid.

This tension is exemplified by Ai2’s change of motto, shifting from “AI for the Common Good” to “Truly Open AI.” The original motto emphasized tangible outcomes and societal benefits, whereas the new mantra treats open-source practices as an end in themselves, divorced from practical consequences. While it would be overreaching to label open source inherently malevolent, the detachment of developers from the real-world impacts of their work evokes Hannah Arendt’s concept of the “banality of evil.” This concept describes how ordinary individuals can perpetrate harm not because they are inherently evil but because of thoughtlessness and blind adherence to orders or conformity to societal norms. In the realm of open-source GenAI, prevailing norms that prioritize minimal restrictions on dataset and model use, along with rapid release cycles aimed at competing with industry giants like Google, Meta, OpenAI, Microsoft, and Anthropic, raise serious ethical concerns that demand careful scrutiny.

The discussion of deployment and release has underscored the ethical debt and societal risks associated with prioritizing open-source GenAI over responsible development. Given these concerns, it is essential to consider how exemptions might be structured to support legitimate scientific research without undermining legal and ethical standards.

# IX. Exemption

Nothing in this paper should be read to mean that activities ordinarily prohibited by law should never be exempted. Instead, exemptions should apply only to open-source initiatives genuinely dedicated to scientific research. To be sure, no safe harbor for data is without weaknesses. Implementation and enforcement can be particularly challenging. However, the alternative to good policy in a space that has the potential to destroy or significantly alter markets, industries, and expectations of privacy cannot be no policy at all.

To qualify for this safe harbor, an institution must meet four conditions: (i) it must be a nonprofit organization; (ii) its primary purpose must be scientific research rather than product development; (iii) all artifacts released must be for non-commercial use and distributed under share-alike licenses; and (iv) access to these artifacts must be restricted to verified, legitimate researchers. Alternatively, groups that do not meet these conditions may seek an exemption from an appropriate democratic body by justifying their need. Such safeguards would not unduly constrain fundamental research; rather, they would reinforce legal frameworks that encourage intellectual property rights, facilitate robust intellectual exchange, and stimulate genuine economic growth. An exemption is unnecessary when GenAI artifacts do not present legal issues. For instance, if a dataset consists exclusively of licensed or public domain materials, the artifacts can be released under any conditions the institution prefers without invoking special exemptions.

Furthermore, responsibility for compliance should extend beyond the developers or distributors of GenAI artifacts. Hosting platforms must also be held accountable for ensuring that the materials they make available adhere to legal and ethical standards. Services like Hugging Face should require comprehensive model cards and datasheets for every GenAI model hosted on their platforms. In addition, they should require that any uploaded content with improper licensing be re-licensed to comply with established guidelines. If an uploader fails to meet these requirements and the platform has provided the uploader with an opportunity to address the issue, the platform must have procedures to remove the content. It is imperative that companies not be allowed to profit from hosting content if they know, or reasonably should know, that such content facilitates breaches of contract, copyright infringement, invasions of privacy, or other legal violations.

Policymakers must be vigilant about a close cousin of openwashing when considering legal exemptions: responsible AI-washing, or RAI-washing. Perhaps the clearest example is Microsoft, which has the open-source Phi series of models. Because there is no democratically designated set of principles for what constitutes RAI, Microsoft can establish its own favorable definition of RAI. It includes six principles: fairness, reliability and safety, privacy and security, inclusiveness, transparency, and accountability.[192]

Interestingly, the transparency principle does not apply to training data, model architecture, or similar aspects. Rather, it focuses on how understandable a model is. Also notable is the accountability principle, which applies to "people"—ostensibly the users of AI systems—not to Microsoft, the developer and deployer of such systems.

But Microsoft's principles are also notable for what they leave out. There is no mention of respecting copyright law, contract law, or any other laws. They also remain silent on their environmental impact or any intention to be transparent about it. Perhaps the least surprising aspect is the silence regarding the violent use of Microsoft's systems: they have no prohibitions, limitations, or commitments to reporting on the use of Microsoft's AI for immigration enforcement, law enforcement, or military purposes. Their "2025 Transparency Report" is no more enlightening[193] Some might be inclined to categorize such oversights as irresponsible.

Google also has a Responsible AI page[194], and it is similarly selective about its commitment to "principles." Earlier this year, it removed a commitment not to pursue:

---

[192] Microsoft, *Responsible AI Principles and Approach*, https://www.microsoft.com/en-us/ai/principles-and-approach (last visited July 30, 2025)

[193] Microsoft, *Our 2025 Responsible AI Transparency Report*, Microsoft On the Issues (June 20, 2025), https://blogs.microsoft.com/on-the-issues/2025/06/20/our-2025-responsible-ai-transparency-report/.

[194] Google, *Google Responsible AI*, https://research.google/teams/responsible-ai/ (last visited July 7, 2025).

- "technologies that cause or are likely to cause overall harm"
- "weapons or other technologies whose principal purpose or implementation is to cause or directly facilitate injury to people"
- "technologies that gather or use information for surveillance violating internationally accepted norms"
- "technologies whose purpose contravenes widely accepted principles of international law and human rights"[195]

The recent attempt by Big Tech and other AI companies to have Congress implement a ten-year moratorium on state AI laws,[196] which had little support from the general population,[197] is yet another example of how RAI, as implemented by AI companies, and democracy are not necessarily aligned.

By proposing a framework for exemptions, this paper aims to balance the need for scientific progress with the imperative of legal and ethical accountability. However, implementing such exemptions would have significant consequences for the future of open-source GenAI. The following section examines these potential outcomes and their broader impact on innovation, competition, and societal trust.

# X. Consequences

If this argument holds, open-source GenAI development could become prohibitively expensive and complex. Yet the purported benefits of open-source GenAI have not been convincingly shown to outweigh the attendant risks. Cybersecurity vulnerabilities, disinformation, misinformation, deepfakes, and similar hazards remain pressing concerns. Unless proponents can provide compelling, concrete evidence that GenAI's societal benefits outweigh these risks, the case for broad legal exemptions remains unconvincing.

Moreover, claims that GenAI is indispensable for national security or victory in the "AI race" are unfounded.[198] The argument for open-source platforms, often touted as a means for the United States to maintain its competitive edge, is equally problematic. For example, Ai2's CEO, Ali Farhadi, claimed that "'If the U.S. wants to maintain its edge … we have only one way, and that is to promote open approaches, promote open-source solutions,' Farhadi added. 'Because no matter how many dollars you're investing in an ecosystem, without communal, global efforts, you're not going to be as fast.'" However, this rationale overlooks a flaw: open-source access invariably extends to all players, including strategic competitors such as China. With equal footing comes an erosion of any claimed competitive advantage, thereby rendering the open-source argument less convincing.[199]

---

[195] Paresh Dave & Caroline Haskins, *Google Lifts a Ban on Using Its AI for Weapons and Surveillance*, WIRED (Feb. 4, 2025), https://www.wired.com/story/google-responsible-ai-principles/.

[196] Cecilia Kang, *Defeat of a 10-Year Ban on State A.I. Laws Is a Blow to Tech Industry*, N.Y. TIMES (July 1, 2025), https://www.nytimes.com/2025/07/01/us/politics/state-ai-laws.html.

[197] Colleen McClain et al., *Views of Risks, Opportunities, and Regulation of AI*, PEW RESEARCH CTR. (Apr. 3, 2025), https://www.pewresearch.org/internet/2025/04/03/views-of-risks-opportunities-and-regulation-of-ai/.

[198] https://www.linkedin.com/posts/scaleai_our-ad-in-the-the-washington-post-january-activity-7287456169084796928-zhD0/ ("America Must Win the AI War"); Emma Roth, *OpenAI and Google ask the government to let them train AI on content they don't own*, The Verge (Mar. 14, 2025), https://www.theverge.com/news/630079/openai-google-copyright-fair-use-exception ("We propose a copyright strategy that would extend the system's role into the Intelligence Age by protecting the rights and interests of content creators while also protecting America's AI leadership and national security. The federal government can both secure Americans' freedom to learn from AI, and avoid forfeiting our AI lead to the PRC by preserving American AI models' ability to learn from copyrighted material.")

Even if the benefits of open-source GenAI are exceptional, we must ask whether these gains truly justify compromising the rights and creative incentives of content creators. We must not, in essence, privilege the interests of technology companies over the legitimate rights of individual contributors.

The analysis of consequences makes clear that open-source GenAI, while promising in some respects, is not a panacea and may introduce new risks and complexities. Ultimately, the case for special legal treatment remains unconvincing unless open-source initiatives can demonstrate clear, tangible benefits that outweigh the harms.

# XI. Conclusion

Legal exemptions for open-source endeavors should be limited to projects whose artifacts are developed for scientific research. While I support responsible open-source innovation, granting companies carte blanche simply because their artifacts are open is legally unsound and ethically questionable.

The principles for responsible open-source GenAI development are straightforward. First, entities must avoid unlawful conduct, including copyright infringement, breaches of terms of service, and violations of privacy norms. Second, developers should commit to producing unbiased, safe, and beneficial datasets and models, actively working to eliminate toxicity or harm and being mindful of environmental impact. Third, they must meet most of the fourteen criteria for being fully open. Put simply, if ethical GenAI excellence cannot be achieved, its pursuit is not justified; the benefits must clearly outweigh the costs.

Moreover, merely increasing access to GenAI components through open-source releases is insufficient to democratize the technology's benefits. The more enduring solution is to empower society through an improved education system. Only an educated populace can fully harness the potential of open GenAI.

In conclusion, open-source GenAI is neither inherently harmful nor a universal remedy. Its benefits, while real, do not automatically outweigh those of proprietary systems, nor do they justify blanket legal exemptions. Only a balanced approach grounded in ethical development, robust oversight, and a commitment to the public interest can ensure that GenAI fulfills its transformative potential without sacrificing legal and societal safeguards.

[199] Ali Farhadi, *Open Source Will Win: Allen Institute for AI CEO Ali Farhadi on the New Era of Artificial Intelligence*, GeekWire (Jan. 22, 2025), https://www.geekwire.com/2025/open-source-will-win-allen-institute-for-ai-ceo-ali-farhadi-on-the-new-era-of-artificial-intelligence/.